\documentclass[a4paper,preprint,superscriptaddress,nofootinbib]{revtex4-1}
\usepackage{graphicx}[dvipdfmx]
\usepackage{bm}
\usepackage{amssymb}
\usepackage{amsmath}
\usepackage{color}
\usepackage{cancel}
\usepackage{comment}
\usepackage{here}

\begin{document} 

\title{Electroweak baryogenesis in aligned two Higgs doublet models}

\author{Kazuki Enomoto}
\email{kenomoto@het.phys.sci.osaka-u.ac.jp}
\affiliation{Department of Physics, Osaka University, Toyonaka, Osaka 560-0043, Japan}

\author{Shinya Kanemura}
\email{kanemu@het.phys.sci.osaka-u.ac.jp }
\affiliation{Department of Physics, Osaka University, Toyonaka, Osaka 560-0043, Japan}

\author{Yushi Mura}
\email{y\_mura@het.phys.sci.osaka-u.ac.jp }
\affiliation{Department of Physics, Osaka University, Toyonaka, Osaka 560-0043, Japan}


\preprint{OU-HET-1114}

\begin{abstract}
We evaluate the baryon number abundance based on the charge transport scenario of top quarks in the CP-violating two Higgs doublet model, in which Yukawa interactions are aligned to avoid dangerous flavor changing neutral currents, and coupling constants of the lightest Higgs boson with the mass $125~\mathrm{GeV}$ coincide with those in the standard model at tree level to satisfy the current LHC data. 
In this model, the severe constraint from the electric dipole moment of electrons, which are normally difficult to be satisfied, can be avoided by destructive interferences between CP-violating phases in Yukawa interactions and scalar couplings in the Higgs potential. 
Viable benchmark scenarios are proposed under the current available data and basic theoretical bounds. 
We find that the observed baryon number can be reproduced in this model, where masses of additional Higgs bosons are typically $300$--$400~\mathrm{GeV}$. 
Furthermore, it is found that the triple Higgs boson coupling is predicted to be $35$--$55~\%$ larger than the standard model value.  
\end{abstract}

\maketitle

\section{Introduction}

Baryon Asymmetry of the Universe (BAU) is one of the remaining big questions of particle physics and cosmology. 
The observed baryon asymmetry from the Big Bang Nucleosynthesis is given by 
\begin{equation}
\label{obsbaryon}
\eta_B^{\mathrm{BBN}} \equiv \frac{n_B-n_{\overline{B}}}{s} = (8.2\text{--}9.2) \times 10^{-11} \text{ at } 95~\%~\mathrm{C.L.},
\end{equation}
where $n_B$ and $n_{\overline{B}}$ are number densities of baryons and anti-baryons, respectively, and $s$ is the entropy density~\cite{ParticleDataGroup:2020ssz}. 
Baryogenesis is one promising interpretation for the origin of the BAU, in which the BAU is created from the baryon symmetric universe after the inflation in the early universe. 
In order to realize baryogenesis, a theory should satisfy the Sakharov conditions; 1) existence of baryon number changing interactions, 2) non-conservation of C and CP, and 3) departure from thermal equilibrium~\cite{Sakharov:1967dj}. 
It has turned out that baryogenesis cannot be realized in the Standard Model (SM)~\cite{EWPT_SM, Huet:1994jb}. 
Therefore, new physics beyond the SM is necessary.
Many scenarios have been proposed for baryogenesis where the Sakharov conditions are satisfied in various mechanisms, such as GUT baryogenesis~\cite{GUT_Baryogenesis}, Affleck-Dine mechanism~\cite{Affleck:1984fy}, Electroweak Baryogenesis (EWBG)~\cite{Kuzmin:1985mm}, Leptogenesis~\cite{Fukugita:1986hr}, et cetera.

In contrast to the other scenarios, EWBG clearly has peculiar importance to be explored at the present stage of particle physics. 
It depends on physics of electroweak symmetry breaking, and the Higgs sector will be throughly tested at various current and future experiments. 
Therefore, the scenario of EWBG can be experimentally confirmed or excluded in the near future.

In the scenario of EWBG, the Sakharov conditions are described  in the following way. 
First, in electroweak theories, baryon number non-conservation is realized by sphaleron transition at high temperatures~\cite{Klinkhamer:1984di}.
Second, C is always violated in chiral electroweak interactions, and CP-violation is provided by interactions in the Higgs sector including Yukawa couplings. 
Third, thermal non-equilibrium is caused by the strongly 1st order electroweak phase transition. 
In the SM, enough CP-violating sources are not supplied~\cite{Huet:1994jb}, and the electroweak phase transition is crossover~\cite{EWPT_SM}, so that some extensions are necessary for a viable scenario of EWBG. 
In fact, although the Higgs boson was discovered, the structure of the Higgs sector and the mechanism of electroweak symmetry breaking remain unknown. 
In addition, the minimal Higgs sector in the SM is tentatively introduced without any theoretical principle. 
There are thus possibilities of extended Higgs sectors, where sufficient additional CP-violating phases are contained and strongly 1st order phase transition is realized, by which viable models of EWBG can be constructed.

Many models for EWBG have been proposed using physics of extended Higgs sectors in the literature. 
The earlier works for the numerical evaluation of the produced baryon number are given in Refs.~\cite{Turok:1990in,Turok:1990zg,Cline:1995dg}. 
Following these works, there have been many studies where the baryon number is evaluated in Two Higgs Doublet Models (THDMs)~\cite{Fromme:2006cm, Cline:2011mm, Tulin:2011wi, Liu:2011jh, Ahmadvand:2013sna, Chiang:2016vgf, Guo:2016ixx, Fuyuto_Senaha, Modak_Senaha, Basler:2021kgq}.\footnote{
There are studies in other models like minimal supersymmetric standard models~\cite{Huet:1995sh, Cline:2000nw, Cirigliano:2006dg}, 
the singlet extension of the standard model~\cite{Espinosa:2011eu, Cline:2012hg}, 
effective field theory approaches~\cite{Bodeker:2004ws, Fromme:2006wx, Cline:2020jre}.}

Physics of CP-violation and that of strongly 1st order electroweak phase transition have been often studied separately in the literature. 
Constraints from the Electric Dipole Moment (EDM) data have been discussed in Refs.~\cite{Pilaftsis:2002fe, Huber:2006ri, Ipek:2013iba, Bian:2014zka, Kanemura:2020ibp}, and physics of the 1st order phase transition have been examined in Refs.~\cite{Turok:1991uc, Anderson:1991zb, Land:1992sm, Hammerschmitt:1994fn, Cline:1996mga, Laine:2000rm, Blinov:2015sna, Inoue:2015pza, Basler:2016obg, Andersen:2017ika, Aoki:2021oez}. 
The prediction on the triple Higgs boson coupling in the scenario of EWBG has been discussed in Ref.~\cite{Kanemura:2004ch}, based on the physics of quantum non-decoupling effects of additional Higgs bosons~\cite{Kanemura:2002vm, Kanemura:2004mg, Braathen_Kanemura}. 
Phenomenological impacts on the di-Higgs production have been studied at hadron, lepton and photon colliders in Refs.~\cite{Pairprod_had1, Pairprod_had2, Goncalves:2018qas, Tian:2013qmi, Kurata:2013, Fujii:2015jha, Asakawa:2010xj} .

Gravitational Waves (GWs) originated from collisions of bubbles of the 1st order electroweak phase transition have also been explored in Refs.~\cite{Apreda:2001us, Grojean:2006bp, Huber:2007vva, Ashoorioon:2009nf, Kakizaki:2015wua, Huang:2015izx}. 
Complementarity between gravitational waves observations and collider experiments to test the 1st order electroweak phase transition has been studied in Ref.~\cite{Hashino:2016rvx, Hashino:2016xoj, Hashino:2018wee}.

In this paper, we evaluate the BAU in the CP-violating THDM in which current EDM constraints are satisfied and strongly 1st order phase transition is realized. 
In order to avoid dangerous Flavor Changing Neutral Currents (FCNCs), we assume alignment in the Yukawa interactions of the model~\cite{Pich:2009sp}. 
In addition, to describe the current constraints on the Higgs couplings at LHC~\cite{CMS:2018uag, ATLAS:2019nkf}, we simply impose the condition that the lightest Higgs boson behaves like the SM one. 
In Ref.~\cite{Aiko:2020ksl}, it has been shown that if this alignment for the Higgs bosons is even slightly broken, most of parameter regions of the model can be explored by the synergy of the direct search at the HL-LHC and the precision measurements of the Higgs boson couplings at the ILC with $\sqrt{s} = 250~\mathrm{GeV}$ with the arguments of perturbative unitarity and vacuum stability, while there are some regions which cannot be excluded in the Higgs aligned scenario, in which this alignment holds exactly at tree level. 
Therefore, in the near future, the Higgs aligned scenario would be more important as the experimental data are accumulated at the collider experiments. 
Constraints on CP-violation in such a scenario from the EDM experiments are studied in this model in Ref.~\cite{Kanemura:2020ibp}, 
and collider phenomenology at proton colliders and $e^+$-$e^-$ colliders are investigated in Ref.~\cite{Kanemura:inpreperation} and Ref.~\cite{Kanemura:2021atq}, respectively. 
In this paper, we evaluate the baryon asymmetry generated by the electroweak baryogenesis in this aligned scenario of the CP-violating THDM. 

We consider the charge transport scenario of top quarks, where the CP-violation via the interaction between top quarks and the Higgs boson is utilized~\cite{Fromme:2006cm, Cline:2011mm}. 
In this framework, the stringent constraint from the electron EDM~\cite{ACME:2018yjb} can be satisfied by the destructive interference of the effects from Yukawa interaction and Higgs potential~\cite{Kanemura:2020ibp}. 
The baryon number is evaluated according to the method developed by Refs.~\cite{Joyce:1994fu, Joyce:1994zn, Cline:2000nw, Fromme:2006wx, Cline:2020jre}. 
We then propose benchmark scenarios of the model where the correct baryon number is evaluated with avoiding the theoretical bounds from perturbative unitarity~\cite{Kanemura:1993hm, Akeroyd:2000wc, Ginzburg:2005dt, Kanemura:2015ska} and vacuum stability~\cite{Nie:1998yn, Kanemura:1999xf, Ferreira:2004yd} and experimental constraints from the current available data from EDM~\cite{ACME:2018yjb, nEDM:2020crw}, LEP~\cite{ALEPH:2013htx}, LHC~\cite{ATLAS:2018rvc, CMS:2019bfg, ATLAS:2020zms, ATLAS:2021upq}, flavor~\cite{CMS:2014xfa, LHCb:2017rmj}. 
Phenomenological consequences for the model can also be discussed. 

What is new in this paper is the following: 
1) focusing on the scenario of the Yukawa alignment and the Higgs alignment, 
2) investigating parameter regions where constraint from electron EDM is avoided by destructive interference of CP-violating effects in the Yukawa coupling and the Higgs potential, and
3) evaluating the BAU in this new scenario.

This paper is organized as follows. 
In Sec.~\ref{sec:model}, we introduce the two Higgs doublet model including the alignments in the Higgs and Yukawa sector which are mentioned above. 
In Sec.~\ref{sec:constraints}, some experimental and theoretical constraints on the model are discussed. 
In Sec.~\ref{sec:effpot}, the effective potentials at one-loop level for zero and finite temperature are shown. 
A formula for the deviation of the triple Higgs boson coupling from that of the SM is also shown. 
In Sec.~\ref{sec:transport}, discussion on EWBG via charge transport by the top quarks is presented. 
In Sec.~\ref{sec:evaluation}, we show the results of numerical evaluations for the baryon asymmetry and the EDMs. 
Some comments on our analysis and phenomenological implication are discussed in Sec.~\ref{sec:discussion},  and conclusions are given in \ref{sec:conclusion}.

\section{The model}
\label{sec:model}
We consider the model including two $SU(2)_L$ doublets $\Phi_1$ and $\Phi_2$, whose hypercharges are $Y=1/2$. 
In general, both doublets can have Vacuum Expectation Values (VEVs) without violating the electromagnetic charge conservation. 
However, by $U(2)$ transformation between the doublets, we can always move into the basis where only one of the doublets has a VEV, so-called the Higgs basis, without loss of generality~\cite{Davidson:2005cw}. 
In the following, we use this basis, and each doublet is parametrized as follows; 
\begin{equation}
\label{eq:Higgs basis}
\Phi_1 = 
\begin{pmatrix}
G^+ \\
\frac{ 1 }{ \sqrt{2} }( v + h_1 + i G^0) \\
\end{pmatrix}
, \quad
\Phi_2 = 
\begin{pmatrix}
H^+ \\
\frac{ 1 }{ \sqrt{2} }( h_2 + i h_3 ) \\
\end{pmatrix}, 
\end{equation}
where $v$ ($= 246~\mathrm{GeV}$) is the VEV. 
The scalars $G^\pm$ and $G^0$ are Nambu-Goldstone bosons which are absorbed into the longitudinal modes of $W^\pm$ and $Z$ boson, respectively. 

In the Higgs basis, the Higgs potential of the model is given by
\begin{align}
\label{eq:scalar_potential}
\mathcal{V} = & -\mu_1^2 |\Phi_1|^2 - \mu_2^2 |\Phi_2|^2 
- \bigl( \mu_3^2 (\Phi_1^\dagger \Phi_2) + \mathrm{h.c.} \bigr)
\nonumber \\
& + \frac{ \lambda_1 }{2 } | \Phi_1 |^4 + \frac{ \lambda_2 }{ 2 } |\Phi_2|^4 
+ \lambda_3 | \Phi_1 |^2 |\Phi_2|^2 
+ \lambda_4 |\Phi_1^\dagger \Phi_2 |^2 
\nonumber \\
& + \biggl\{
	\Bigl( \frac{ \lambda_5 }{ 2 } (\Phi_1^\dagger \Phi_2) 
		+ \lambda_6 | \Phi_1 |^2 + \lambda_7 | \Phi_2 |^2 
	\biggr) (\Phi_1^\dagger \Phi_2 ) + \mathrm{h.c.}
\Bigr\}. 
\end{align}
The coupling constants $\mu_3^2$, $\lambda_5$, $\lambda_6$, and $\lambda_7$ are generally complex numbers, however, one of them can be real by the redefinition of the phase of $\Phi_2$. 

By substituting Eq.~(\ref{eq:Higgs basis}) into the Higgs potential, 
we obtain the following stationary conditions; 
\begin{equation}
\label{eq:stationary_condition}
\mu_1^2 = \frac{ \lambda_1 }{ 2 } v^2 , \quad 
\mu_3^2 = \frac{ \lambda_6 }{ 2 } v^2 .
\end{equation}
Because of the second condition, the CP-violating phases of $\mu_3^2$ and $\lambda_6$ are the same. Therefore, independent CP-violating phases in the Higgs potential are generally two of $\arg[\lambda_5]$, $\arg[\lambda_6]$, and $\arg[\lambda_7]$. 

The charged scalars $H^\pm$ are mass eigenstates without mixing, 
and their mass $m_{H^\pm}^{}$ is given by
\begin{equation}
\label{eq:mass_of_Hpm}
m_{H^\pm}^2 = M^2 + \frac{ \lambda_3 }{ 2 } v^2, 
\end{equation} 
where $M^2 = - \mu_2^2$. 
On the other hand, the neutral scalars $h_1$, $h_2$ and $h_3$ are generally mixed, 
and their mass terms are given by $\frac{ 1 }{ 2 } \sum_{i,j} h_i \mathcal{M}^2_{ij} h_j$, where 
\begin{equation}
\label{eq:mass_matrix}
\mathcal{M}^2 = v^2
\begin{pmatrix}
\lambda_1 & \mathrm{Re}[\lambda_6] & - \mathrm{Im}[\lambda_6] \\
\mathrm{Re}[\lambda_6] & \frac{M^2}{v^2} + \frac{ 1 }{ 2 } \bigl( \lambda_3 + \lambda_4 + \mathrm{Re}[\lambda_5]\bigr) & - \frac{ 1 }{ 2 } \mathrm{Im}[\lambda_5] \\
- \mathrm{Im}[\lambda_6] & - \frac{ 1 }{ 2 } \mathrm{Im}[\lambda_5] & \frac{M^2}{v^2} + \frac{ 1 }{ 2 } \bigl( \lambda_3 + \lambda_4 - \mathrm{Re}[\lambda_5]\bigr) \\
\end{pmatrix}. 
\end{equation}
As mentioned above, $\mathrm{Im}[\lambda_5]$ can be $0$ by the redefinition of the phase of $\Phi_2$. 
Then, the mixings between neutral scalar states at tree level are caused by only one coupling constant~$\lambda_6$. 
For simplicity, we consider this case, thus $\lambda_5$ is a real number. 

Mass eigenstates of neutral scalars ($H_1$, $H_2$, $H_3$) are defined by 
\begin{equation}
H_i = \sum_{j=1}^3 \mathcal{R}_{ji} h_j, 
\end{equation}
where the matrix $\mathcal{R}$ is an orthogonal matrix which diagonalizes the mass matrix as $\mathcal{R}^\mathrm{T} \mathcal{M}^2 \mathcal{R} = \mathrm{diag}(m_{H_1}^2, m_{H_2}^2, m_{H_3}^2)$. 
If the non-diagonal elements $R_{12}$ and $R_{13}$ are nonzero, 
they induce deviations of the Higgs couplings from the SM values at tree level. 
In order to avoid them, in the following, we assume the alignment condition, $\lambda_6 = 0$~\cite{Kanemura:2020ibp}. 
Then, the matrix $\mathcal{R}$ is an identity matrix ($R_{ij} = \delta_{ij}$), 
and the squared mass of each neutral scalar $H_i$ ($i=1,2,3$) is given by the diagonal elements $\mathcal{M}^2_{ii}$: 
\begin{align}
&m_{H_1}^2 = \lambda_1 v^2, \\ 
\label{eq:mass_of_H2}
&m_{H_2}^2 = M^2 + \frac{ v^2 }{ 2 }(\lambda_3 + \lambda_4 + \lambda_5), \\
\label{eq:mass_of_H3}
&m_{H_3}^2 = M^2 + \frac{ v^2 }{ 2 }( \lambda_3 + \lambda_4 - \lambda_5). 
\end{align}
The scalar boson $H_1(=h_1)$ is the SM Higgs boson, and the value of $\lambda_1$ is determined by the mass of the Higgs boson, $m_{H_1} = 125~\mathrm{GeV}$. 
As a result, the real free parameters in the Higgs potential are $M^2(=-\mu_2^2)$, $m_{H^\pm}$, $m_{H_2}$, $m_{H_3}$, 
$\lambda_2$, $|\lambda_7|$, and $\theta_7 (= \arg[\lambda_7])$. 
We note that only $\theta_7$ is the CP-violating parameter in the Higgs potential.

The Yukawa interaction in the model is given by 
\begin{equation}
-\mathcal{L}_Y = \sum_{a=1}^2 \sum_{i,j=1}^3 
\Bigl( \overline{Q_{iL}^\prime} (y_u^a)^\dagger_{ij} \tilde{\Phi}_a u_{jR}^\prime 
	+ \overline{Q_{iL}^\prime} (y_d^a)_{ij} \Phi_a d_{jR}^\prime
	+ \overline{L_{iL}^\prime} (y_e^a)_{ij} \Phi_a e_{jR}^\prime
\Bigr), 
\end{equation}
where $\tilde{\Phi}_a$ are defined by using the Pauli matrix $\sigma_2$ as $\tilde{\Phi}_a = i \sigma_2 \Phi_a^\ast$. 
The fermion $Q_{iL}^\prime$ ($L_{iL}$) is the left-handed isospin doublet of quarks (leptons), 
and $u^\prime_{iR}$, $d_{iR}^\prime$, and $e_{iR}^\prime$ are 
right-handed isospin singlet of up-type quark, down-type quark, and lepton, respectively, 
where $i, j = 1,2,3$ is the indeces of flavors.   
In general, it is not possible to diagonalize both of Yukawa matrices $y_f^1$ and $y_f^2$ ($f=u,d,e$) simultaneously, 
and it causes the FCNCs at tree level~\cite{Glashow:1976nt}. 
In order to avoid it, we impose the Yukawa alignment proposed by Pich and Tuzon~\cite{Pich:2009sp}; 
\begin{equation}
y_f^2 = \zeta_f y_f^1, \quad f=u,d,e, 
\end{equation}
where each $\zeta_f$ is a complex number. 
Then, $y_f^1$ and $y_f^2$ can be diagonalized simultaneously, and there is no FCNCs at tree level as in the SM. 
In the Yukawa sector, there are three CP-violating phases $\theta_f \equiv \mathrm{arg}[\zeta_f]$ ($f=u,d,e$). 

The Yukawa interactions between each scalar mass eigenstate and the SM fermions are given by
\begin{align}
\label{eq:Yukawa_interactions}
-\mathcal{L}_Y \ni \sum_{i, j=1}^3
\biggl\{
	& \sum_{k =1}^3
	\Bigl( 
		\sum_{f=u,d,e}  \frac{m_{f_i}}{v} \delta_{ij} \kappa_f^k \overline{f_{iL}} f_{jR} 
	\Bigr) H_k
	\nonumber \\
	+ & \frac{ \sqrt{2} }{ v } 
	\Bigl( 
		- \zeta_u \overline{u_{iR}} (m_{u_i} V_{ij}) d_{jL}
		+ \zeta_d \overline{u_{iL}} (V_{ij} m_{d_j}) d_{jR}
		+ \zeta_e \overline{\nu_{iL}} m_{e_i} e_{jR} 
	\Bigr) H^+ 
\biggr\}
+ \mathrm{h.c.}, 
\end{align}	
where all fermions are mass eigenstates, and $m_{f_i}$ is the mass of fermion $f_i$. 
The matrix $V$ is the CKM matrix~\cite{Cabibbo:1963yz, Kobayashi:1973fv}.\footnote{
In our analysis, we neglect the Kobayashi-Maskawa CP-violating phase in the CKM matrix because its effect on the EDMs is neglegible.}
The constants $\kappa_f^k$ in Eq.~(\ref{eq:Yukawa_interactions}) are defined by
\begin{equation}
\left\{
\begin{array}{l l l l}
\kappa_f^1 = 1, & \kappa_f^2 = \zeta_f, & \kappa_f^3 = i \kappa_f^2, &  (f=d,e), \\[10pt]
\kappa_u^1 = 1, & \kappa_u^2 = \zeta_u^\ast, & \kappa_u^3 = - i \kappa_u^2. & \\
\end{array}
\right.
\end{equation}
The relation between the Yukawa sector of the model and that of the THDM with softly-broken $Z_2$ symmetry~\cite{Glashow:1976nt, Barger:1989fj, Grossman:1994jb, Aoki:2009ha} is shown in Ref.~\cite{Pich:2009sp, Kanemura:2020ibp}.

\section{Theoretical and experimental constraints}
\label{sec:constraints}
In this section, we consider some experimental and theoretical constraints on the model.  
For simplicity, we discuss only the case where all $\zeta_f$ ($f=u,d,e$) have the same absolute value; $ |\zeta_u| = |\zeta_d| =|\zeta_e|$. In this case, the Yukawa couplings except for CP-violating phases are the same with that in the so-called Type-I THDM~\cite{Barger:1989fj, Grossman:1994jb, Aoki:2009ha}.  

First, we consider the constraints from the direct searches for the charged scalar boson $H^\pm$ according to Refs.~\cite{Arbey:2017gmh, Aiko:2020ksl, Akeroyd:2016ymd}. 
From the LEP results, the lower bound on the mass of charged Higgs boson $H^\pm$ is given by $m_{H^\pm} \gtrsim 80~\mathrm{GeV}$~\cite{ALEPH:2013htx}. 
This bound is almost independent on the value of $|\zeta_f|$. 
As the production processes of $H^\pm$ at the LHC, 
we consider the production from the decay of top quark $t \to H^\pm b$ and the associated production $g g \to H^\pm tb$ ($gb \to t H^\pm$). 
In the mass region $80~\mathrm{GeV} \lesssim m_{H^\pm} \lesssim 170~\mathrm{GeV}$, $H^\pm$ is dominantly produced via $t \to H^\pm b$. 
By using the latest data at CMS~\cite{CMS:2019bfg}, 
the upper bound on $|\zeta_f|$ is estimated to be about $0.07$ ($0.27$) for $m_{H^\pm} = 80~\mathrm{GeV}$ ($160~\mathrm{GeV}$). 
In the mass region $m_{H^\pm} \gtrsim 170~\mathrm{GeV}$, 
the top quarks cannot decay into $H^\pm b$, and $H^\pm$ is dominantly generated via the associated production $g g \to H^\pm tb$. 
By using the latest data at ATLAS~\cite{ATLAS:2021upq}, 
the upper bound on $|\zeta_f|$ is estimated to be about $0.5$ ($0.6$) for $m_{H^\pm} = 200~\mathrm{GeV}$ ($400~\mathrm{GeV}$).  

Second, we consider the constraints from the direct searches for the additional neutral scalar bosons $H_{2,3}$ according to Refs.~\cite{Arbey:2017gmh, Aiko:2020ksl} in the case neglecting the CP-violating phases. 
As the production processes of $H_{2,3}$ at the LHC, 
we consider the single production via the top quark loop $gg \to H_{2,3}$ and the bottom associated production $gg \to H_{2,3} b \overline{b}$ ($b \overline{b} \to H_{2,3}$). 
Since we assume that $ |\zeta_u| = |\zeta_d| =|\zeta_e|$, 
the production via the top quark loop is dominant.
If we consider the case that $|\zeta_d| \gg |\zeta _u|$, 
the bottom associated production is also important. 
For $m_{H_{2,3}} < 2 m_t$ ($m_{H_{2,3}} > 2 m _t$), 
the process $H_{2,3} \to \tau \overline{\tau}$ ($H_{2,3} \to t \overline{t}$) gives a strong constraint on $|\zeta_f|$. 
By using the result of the search for $H_{2,3} \to \tau \overline{\tau}$ in Ref.~\cite{ATLAS:2020zms}, 
the upper bound on $|\zeta_f|$ is estimated to be about $0.35$ for $m_{H_3} = 200~\mathrm{GeV}$. 
On the other hand, 
the upper bound from $H_{2,3} \to t \overline{t}$ is estimated to be about $0.67$ for $m_{H_3} = 400~\mathrm{GeV}$ by using the result in Ref.~\cite{ATLAS:2018rvc}. 

Third, we consider the constraints from the flavor experiments. 
The constraints on the charged Higgs boson from various flavor experiments in softly $Z_2$ symmetric THDMs are investigated in Refs.~\cite{Enomoto:2015wbn, Arbey:2017gmh, Haller:2018nnx}. 
According to Refs~\cite{Haller:2018nnx}, 
under the assumption $ |\zeta_u| = |\zeta_d| =|\zeta_e|$ which corresponds to Type-I THDM, 
the most stringent constraint is given by $B_d \to \mu \mu$~\cite{CMS:2014xfa, LHCb:2017rmj}. 
The bound on $|\zeta_f|$ depends on the mass of $H^\pm$, 
and it is $|\zeta_f| \lesssim 0.33$ for $m_{H^\pm} = 100~\mathrm{GeV}$ 
and $|\zeta_f| \lesssim 0.5$ for $m_{H^\pm} = 400~\mathrm{GeV}$ at $95~\%$ C.L. 

\begin{figure}[t]
	\begin{minipage}[t]{0.45\textwidth}
	\centering
	\includegraphics[width = 50mm]{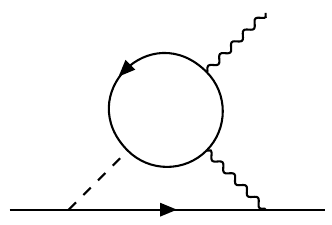} 
   	 \caption{A Barr-Zee type diagram with a fermion loop}
   	 \label{fig:BZf}
	 \end{minipage}
	 \hspace{15pt}
	 \begin{minipage}[t]{0.45\textwidth}
	 \centering
	 \includegraphics[width = 50mm]{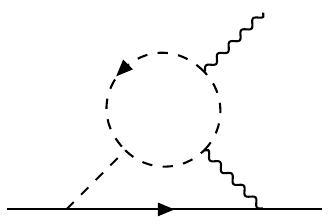} 
 	 \caption{A Barr-Zee type diagram with a scalar loop}
 	 \label{fig:BZs}
	 \end{minipage}
\end{figure}

Fourth, we consider the constraints from the EDM experiments. 
The EDM of a fermion $f$, $d_f$ is given by the coefficient of the following dimension five operators; 
\begin{equation}
\mathcal{L}_\mathrm{EDM} = - \frac{ d_f }{ 2 } \overline{f} \sigma^{\mu\nu} (i\gamma_5) f F_{\mu \nu}, 
\end{equation}
where $F_{\mu\nu}$ is the field strength of the electromagnetic field. 
From the ACME experiment~\cite{ACME:2018yjb}, 
the constraint from electric EDM (eEDM) is given by $|d_e + k C_s | < 1.1 \times 10^{-29}~\mathrm{e\, cm}$ at $90~\%$ C.L., 
where $C_s$ is the coefficient of the dimension six interaction between electrons and nucleons $C_s (\overline{e} i \gamma_5 e)(\overline{N} N)$, 
and the constant $k$ is about $\mathcal{O}(10^{-15})~\mathrm{GeV^2\  e\, cm}$. 
We checked that 
as discussed in Refs.~\cite{Jung:2013hka, Cheung:2014oaa, Kanemura:2020ibp}, 
in the parameter region where we discuss later, 
the typical value of the second term is two orders smaller than the current constraint. 
Consequently, as the constraint from the ACME experiment, 
we require that $|d_e| < 1.0 \times 10^{-29}~\mathrm{e\, cm}$. 

In the model, new scalars contribute to $d_e$ via 2-loop Barr-Zee type diagrams~\cite{Barr:1990vd}.  
There are two kinds of diagrams shown in Figs.~\ref{fig:BZf} and~\ref{fig:BZs}. 
In Fig.~\ref{fig:BZf} (Fig.~\ref{fig:BZs}), the diagram including a fermion loop (scalar loop) is shown. 
Since we assume that $ |\zeta_u| = |\zeta_d| =|\zeta_e|$, 
the largest contribution of fermion loop diagrams is given by the diagram including the top quark loop, and values of the other fermion loop diagrams are negligibly small. 
As a result, fermion loop diagrams are predominantly proportional to $\sin (\theta_u - \theta_e)$, 
while scalar loop diagrams are proportional to $\sin (\theta_7 - \theta_e)$. 
By using the destructive interference between $\theta_7$ and $\theta_u$, the current eEDM constraint can be satisfied with keeping each CP-violating phase to be $\mathcal{O}(1)$~\cite{Kanemura:2020ibp}. 

Latest constrains from the neutron EDM (nEDM) is given by the NEDM collaboration as $|d_n| < 1.8 \times 10^{-26}~\mathrm{e\, cm}$ at $90~\%$ C.L~\cite{nEDM:2020crw}. 
By using QCD sum rule, $d_n$ can be evaluated by
\begin{equation}
\label{eq:nEDM}
d_n = 0.79 d_d - 0.20 d_u + e (0.59 d_d^C + 0.30 d_u^C) / g_3, 
\end{equation}
where $g_3$ is the strong coupling constant, and $d_q^C$ ($q=u,d$) is chromo EDM~\cite{Abe:2013qla, Pospelov:2000bw,  Hisano:2012sc, Fuyuto:2013gla}. 
Other operators such as the Weinberg operator~\cite{Weinberg:1989dx, Dicus:1989va} and four Fermi interactions~\cite{Khatsimovsky:1987fr}, also contribute to the nEDM. 
In the parameter region where we discuss later, 
the contribution from the Weinberg operator is as large as that from Eq.~(\ref{eq:nEDM}), while that from four Fermi interactions is negligibly small~\cite{Jung:2013hka}.
Therefore, we consider only the contributions from $d_n$ and the Weinberg operator.\footnote{In considering the contribution from the Weinberg operator, 
we take its absolute value because the sign of it includes theoretical uncertainties~\cite{Jung:2013hka}.}
Because of the assumption $ |\zeta_u| = |\zeta_d| =|\zeta_e|$, 
the nEDM is predominantly proportional to $\sin (\theta_u - \theta_d)$. 

Fifth, we discuss the oblique parameters $S$, $T$, and $U$~\cite{oblique_parameter}, 
especially the $T$ parameter. 
In the scalar potential, $\lambda_4 - \lambda_5$ and $\mathrm{Im}[\lambda_7]$ violate the custodial symmetry~\cite{Sikivie:1980hm, Haber:1992py, Pomarol:1993mu, Gerard:2007kn, Haber:2010bw, Grzadkowski:2010dj, Aiko:2020atr}, and they induce the deviation of the $T$ parameter from the SM prediction. 
Since $\lambda_4 - \lambda_5$ is proportional to $m_{H_3}^2 - m_{H^\pm}^2$, it vanishes when the masses of $H^\pm$ and $H_3$ are degenerated. 
In addition, as shown in Ref~\cite{Haber:2010bw, Pomarol:1993mu}, $\mathrm{Im}[\lambda_7]$ does not contribute to the $T$ parameter at one-loop level. 
Therefore, when we assume that $m_{H_3} = m_{H^\pm}$, 
new scalars do not contribute to the $T$ parameter at one-loop level. 

Finally, the extended Higgs sector is constrained theoretically by the conditions from perturbative unitarity~\cite{Kanemura:1993hm, Akeroyd:2000wc, Ginzburg:2005dt, Kanemura:2015ska} and vacuum stability~\cite{Nie:1998yn, Kanemura:1999xf, Ferreira:2004yd}.  
In our analysis, we employ these conditions given in Refs.~\cite{Kanemura:2015ska} and~\cite{Ferreira:2004yd}.

\section{The effective potential}
\label{sec:effpot}
In this section, we consider the effective potential of the model. 
In the calculation, the neutral elements of $\Phi_1$ and $\Phi_2$ are shifted by $\frac{ 1 }{ \sqrt{2} } \varphi_1$ and $\frac{ 1 }{ \sqrt{2} } (\varphi_2 + i \varphi_3)$, respectively. Each $\varphi_i$ ($i=1,2,3$) is a real constant field, 
and we do not consider the imaginary part of the neutral elements of $\Phi_1$ 
since it can be $0$ by an appropriate $SU(2)_L$ transformation. 

At one-loop level, the effective potential at zero temperature is given by 
\begin{equation}
V_{T=0}(\varphi_1, \varphi_2, \varphi_3) = V_0 + V_1 + V_{CT}, 
\end{equation}
where $V_0$ is the tree level potential which is given in Eq.~(\ref{eq:scalar_potential}), $V_1$ is Coleman-Weinberg potential which is at one-loop level~\cite{Coleman:1973jx}, and $V_{CT}$ is the counterterms. 
The effective potential is a gauge-dependent quantity~\cite{Jackiw:1974cv}.  
We calculate loop diagrams in the Landau gauge. 
Then, top quark $t$, weak bosons $W^\pm$ and $Z$, photon $\gamma$, and the scalar bosons $H^\pm$, and $H_{1,2,3}$ contribute to the one-loop diagrams,\footnote{We neglect the contribution from other SM fermions because they are enough small in the case that $|\zeta_u| = |\zeta_d| = |\zeta_e|$.} and $V_1$ is given by 
\begin{equation}
V_1 = \sum_k (-1)^{s_k} \frac{ n_k }{ 64 \pi^2 } \tilde{m}_k^4
\biggl[
	\log \frac{ \tilde{m}_k^2 }{ Q^2 } - \frac{ 3 }{ 2 } 
\biggr], 
\end{equation}
where $k$ is the label of the kind of particles in internal lines; 
$k=t, W^\pm, Z, \gamma, H^\pm, H_1, H_2, H_3$. 
The constant $s_k$ is $0$ ($1$) if $k$ is a scalar (fermion), 
and $n_k$ is the degree of freedom for $k$. 
The symbol $\tilde{m}_k$ is the field dependent mass of $k$ and is a function of $\varphi_1$, $\varphi_2$, and $\varphi_3$, 
and $Q$ denotes the renormalization scale which is assumed to be the mass of the $Z$ boson in our numerical evaluation in Sec.~\ref{sec:evaluation}. 

In order to determine the counterterms, 
we impose the following conditions, 
\begin{align}
& \left. \frac{ \partial V_{T=0} }{ \partial \varphi_i } \right|_{\substack{\varphi_1 = v,\\  \varphi_2 = \varphi_3 = 0}} = 0, \quad (i=1,2,3), \\
& \left. \frac{ \partial^2 V_{T=0} }{ \partial \varphi_i \partial \varphi_j }  \right|_{\substack{\varphi_1 = v,\\  \varphi_2 = \varphi_3 = 0}} = \mathcal{M}^2_{ij}, \quad (i,j=1,2,3), 
\end{align}
where $\mathcal{M}^2$ is the mass matrix defined in Eq.~(\ref{eq:mass_matrix}). 
The first condition determines the vacuum and the second condition gives six independent equations. 
By using these conditions, parameters in $V_{CT}$ except for those of $\mu_2^2$, $\lambda_2$, and $\lambda_7$ can be determined. 
In order to fix left three counterterms, we employ the $\overline{\text{MS}}$ scheme. 
In the calculation of the second derivative in the renormalization conditions, 
NG bosons cause IR divergences, therefore, we use the IR cutoff regularization where the mass of NG bosons at the vacuum $m_{NG}$ is replaced by the IR cutoff scale $1~\mathrm{GeV}$~\cite{Baum:2020vfl}.

By using the effective potential at zero temperature, 
the triple Higgs boson coupling $\lambda_{hhh}$ can be evaluated by \cite{Kanemura:2002vm, Kanemura:2004mg}
\begin{equation}
\lambda_{hhh} = \left. \frac{ \partial^3 V_{T=0} }{ \partial \varphi_1^3 } \right|_{\substack{\varphi_1 = v,\\  \varphi_2 = \varphi_3 = 0}}. 
\end{equation}
The deviation in $\lambda_{hhh}$ from the SM prediction is defined as $\Delta R = (\lambda_{hhh} - \lambda_{hhh}^{SM})/\lambda_{hhh}^{SM}$, 
and $\Delta R$ in the model at one-loop level is given by
\begin{equation}
\label{eq:DeltaR}
\Delta R = \frac{ 1 }{ 12 \pi^2 v^2 m_{H_1}^2 } 
\biggl\{
	2 \frac{(m_{H^\pm}^2 - M^2)^3 }{ m_{H^\pm}^2 }
	+ \frac{ (m_{H_2}^2 - M^2)^3 }{ m_{H_2}^2 }
	+ \frac{ (m_{H_3}^2 - M^2)^3 }{ m_{H_3}^2 }
\biggr\}, 
\end{equation}
in the Higgs alignment limit.

At finite temperature, 
the effective potential includes the term from thermal effects in addition to $V_{T=0}$;
\begin{equation}
V(\varphi_1, \varphi_2, \varphi_3; T) = V_{T=0} + V_T, 
\end{equation} 
where $V_T$ denotes the thermal effects, and it is given by 
\begin{equation}
V_T = \sum_k (-1)^{s_k} \frac{ n_k }{ 2 \pi^2 \beta^4 } \int_0^\infty \mathrm{d}x \, 
x^2 \log\biggl( 1 + (-1)^{s_k+1} \exp \Bigl( - \sqrt{x^2 + \beta^2\tilde{m}^2_k  } \Bigr) \biggr), 
\end{equation}
where $\beta = 1 / T$~\cite{Dolan:1973qd}. 
For thermal resummation, we use the Parwani scheme, where the field dependent masses $\tilde{m}^2_k$ in $V_1$ and $V_T$ are replaced by the masses including thermal effects~\cite{Parwani:1991gq}. 

\section{Electroweak baryogenesis in the model}
\label{sec:transport}
In this section, we consider the baryon asymmetry generated via the charge transport by top quarks at the electroweak phase transition. 
The charge transport by the top quarks is investigated in the THDM in Refs.~\cite{Fromme:2006cm, Cline:2011mm}.
We consider the same situation in these references. 

We assume that the velocity of the bubble wall $v_w$ is small, and we use the linear expansions about $v_w$. 
In the following, the problem is discussed in the wall frame, where the bubble wall is stationary. 
The radial coordinate in the wall frame which is perpendicular to the surface of the bubble is denoted by $z$. 
The bubble wall is at $z=0$, and the negative direction of $z$ is defined as pointing toward the center of the bubble (broken phase). 

First, we derive transport equations for the chemical potentials of each particle from the Boltzmann equation in the wall frame.   
We here assume that the baryon number is conserved, and the weak sphaleron is not included. 
In order to derive transport equations, we use the WKB approximation method~\cite{Joyce:1994fu, Joyce:1994zn, Cline:2000nw, Fromme:2006wx, Cline:2020jre}. 
In addition, we neglect the mass of the bottom quark and Higgs boson, because their effects are enough small according to Ref.~\cite{Cline:2020jre}.

In the transport equations, effects of the strong sphaleron process~\cite{McLerran:1990de, Giudice:1993bb}, $W$-scattering, the top Yukawa interaction, the top helicity flips, and the Higgs number violation are included~\cite{Fromme:2006wx, Fromme:2006cm}, and the rates of each process are denoted by $\Gamma_{ss}$, $\Gamma_W$, $\Gamma_y$, $\Gamma_m$, and $\Gamma_h$, respectively. 
The quarks in the first and second generations and the right-handed bottom quarks are included in only the strong sphaleron process. 
Then, the transport equation for the quarks in the first and second generations and the right-handed bottom quarks can be solved analytically, and the chemical potentials for them can be represented as functions of those of top quarks and left-handed bottom quarks. 
By using these solutions, the transport equations for top quarks, left-handed bottom quarks, and Higgs bosons are given as follows~\cite{Fromme:2006wx, Fromme:2006cm, Cline:2020jre}. 
\begin{itemize}
\item Left-handed top quarks ($t$)
\begin{align}
\label{eq:first_transport_eq_for_tL}
& v_w K_{1,t} \mu_t^\prime + v_w K_{2,t} (M_t^2)^\prime \mu_t + u_t^\prime - K_{0,t} \overline{\Gamma}_t = 0, \\
\label{eq:second_transport_eq_for_tL}
& - K_{4,t} \mu_t^\prime - v_w u_t^\prime + v_w \tilde{K}_{6,t} (M_t^2)^\prime u_t + \Gamma_\mathrm{tot}^t u_t + v_w K_{0,t} \overline{\Gamma}_t  = - S_t. 
\end{align}
\item Charge conjugation of right-hand top quarks ($t^c$)
\begin{align}
\label{eq:first_transport_eq_for_tR}
& v_w K_{1,t} \mu_{t^c}^\prime + v_w K_{2,t} (M_t^2)^\prime \mu_{t^c} + u_{t^c}^\prime - K_{0,t} \overline{\Gamma}_{t^c} = 0, \\
\label{eq:second_transport_eq_for_tR}
& - K_{4,t} \mu_{t^c}^\prime - v_w u_{t^c}^\prime + v_w \tilde{K}_{6,t} (M_t^2)^\prime u_{t^c} + \Gamma_\mathrm{tot}^t u_{t^c} + v_w K_{0,t} \overline{\Gamma}_{t^c}  = - S_t. 
\end{align}
\item left-hand bottom quarks ($b$)
\begin{align}
& v_w K_{1,b} \mu_b^\prime + u_b^\prime - K_{0,b} \overline{\Gamma}_b = 0, \\
& - K_{4,b} \mu_b^\prime - v_w u_b^\prime + \Gamma_\mathrm{tot}^b u_b + v_w K_{0,b} \overline{\Gamma}_b = 0, 
\end{align}
\item Higgs ($h$)
\begin{align}
&v_w K_{1,h} \mu_h^\prime + u_h^\prime - K_{0,h} \overline{\Gamma}_h = 0, \\
\label{eq:second_transport_eq_for_Higgs}
&- K_{4,h} \mu_h^\prime - v_w u_h^\prime + \Gamma_\mathrm{tot}^h u_h + v_w K_{0,h} \overline{\Gamma}_h = 0,  
\end{align}
\end{itemize} 
where $S_t$ in Eqs.~(\ref{eq:second_transport_eq_for_tL}) and (\ref{eq:second_transport_eq_for_tR}) is defined as 
\begin{equation}
\label{eq:source_term}
S_t = - v_w K_{8,t}(M_t^2 \theta_t^\prime)^\prime 
+ v_w K_{9,t} \theta^\prime M_t^2 (M_t^2)^\prime, 
\end{equation}
where the functions $M_t$ and $\theta_t$ are the absolute value and the phase of the localized top quark mass. 
They are calculated by using solutions of the bounce equations for $\varphi_1$, $\varphi_2$, and $\varphi_3$.  
We define the functions $\hat{\varphi}_i(z)$ ($i=1,2,3$) as the solutions of the bounce equations for $\varphi_i$. 
Then, $M_t$ and $\theta_t$ are given by~\cite{Cline:2011mm}
\begin{align}
&M_t(z) = \frac{ m_t }{ v }  
\bigl( 
	\hat{\varphi}_1^2 + 2 |\zeta_u| \hat{\varphi}_1 \hat{\varphi}_H \cos ( \theta_H + \theta_u) + |\zeta_u|^2 \hat{\varphi}_H^2 
\bigr)^{1/2}, \\
&\partial_z \theta_t(z) = - \frac{ \hat{\varphi}_H^2 }{ \hat{\varphi}_1^2 + \hat{\varphi}_H^2 }
						\partial_z \theta_H
						+ \partial_z \mathrm{Tan}^{-1}
						\left( 
							\frac{ \hat{\varphi}_H |\zeta_u|\sin(\theta_H + \theta_u) }{ \hat{\varphi}_1 + \hat{\varphi}_H |\zeta_u| \cos (\theta_H + \theta_u) }
						\right), 
\end{align}
where $\hat{\varphi}_H(z) = \sqrt{ \hat{\varphi}_2^2 + \hat{\varphi}_3^2}$ and $\theta_H (z) = \mathrm{Tan}^{-1}(\hat{\varphi}_3/\hat{\varphi}_2)$. 

We explain the meanings of undefined parameters in Eqs.~(\ref{eq:first_transport_eq_for_tL})-(\ref{eq:source_term}) in order.  
The symbol $\mu_i$ ($u_i$) for $i=t$, $t^c$, $b$, and $h$ is CP-odd components of the local chemical potential (the plasma velocity) of the particle $i$, which is a function of the radial coordinate $z$. 
The prime in the equations means the derivative by $z$. 
The functions given by $K_{a,i}$ ($a = 0$--$9$, $i=t,b,h$) are defined in Ref.~\cite{Fromme:2006wx}.\footnote{
Actually, formulae in Ref.~\cite{Fromme:2006wx} include some errors, for example, in the definition of $\tilde{K}_{6,i}$, which are mentioned in Ref.~\cite{Cline:2020jre}. 
In our analyses, these errors have been collected, and our formulae are the same as those in Ref.~\cite{Cline:2020jre} at the linear order of $v_w$. }
The symbol $\Gamma_\mathrm{tot}^i$ is the rate of the total scattering for $i$, 
and $\overline{\Gamma}_i$ denotes the effect from the inelastic scatterings for $i$ which is given by
\begin{align}
&\overline{\Gamma}_{t} = \Gamma_{ss} \Bigl( ( 1 + 9 K_{1,t} ) \mu_{t_L} + ( 1 + 9 K_{1,b} ) \mu_{b} + ( 1 - 9 K_{1,t} )\mu_{t^c} \Bigr)
\nonumber \\
& \hspace{20pt} +\Gamma_W (\mu_{t} - \mu_{b} )
+\Gamma_y ( \mu_{t} + \mu_{h} + \mu_{t^c} )
+ \Gamma_m ( \mu_{t} + \mu_{t^c}), 
\\
&\overline{\Gamma}_{t^c} = \Gamma_{ss} \Bigl( ( 1 - K_{1,t} )\mu_{t^c} + ( 1 + 9 K_{1,t} ) \mu_{t} + ( 1 + 9 K_{1,b} ) \mu_{b} \Bigr)
\nonumber \\
& \hspace{20pt} + \Gamma_m ( \mu_{t^c} + \mu_{t} )
+ \Gamma_y ( 2 \mu_{t^c} + \mu_{t} + \mu_{b} + 2 \mu_h ) , 
\\ 
& \overline{\Gamma}_b = \Gamma_{ss} \bigl( ( 1 + 9 K_{1,t} ) \mu_{t} + ( 1 + 9 K_{1,b} ) \mu{b} + ( 1 - 9 K_{1,t} )\mu_{t^c} \bigr)
\nonumber \\
& \hspace{20pt} + \Gamma_W (\mu_{b} - \mu_{t} )
+\Gamma_y (  \mu_{b} + \mu_h +  \mu_{t^c} ), 
\\
& \overline{\Gamma}_h = \Gamma_y (  2 \mu_h + \mu_{t} + \mu_{b} + 2 \mu_{t^c} ) 
+ \Gamma_h \mu_h.
\end{align} 

By solving the transport equations in Eqs.~(\ref{eq:first_transport_eq_for_tL})-(\ref{eq:second_transport_eq_for_Higgs}) with the boundary conditions $\mu_i (z=\pm \infty) = 0$ ($i=t,t^c,b,h$), 
we can obtain the distributions of the CP-odd chemical potentials for each particle. 
Then, we can calculate the baryon asymmetry $\eta_B$ normalized by the entropy density $s$ by the following formula~\cite{Cline:2000nw, Cline:2011mm}, 
\begin{equation}
\label{eq:baryon_asymmetry}
\eta_B = \frac{ 405 \Gamma_\mathrm{sph} }{ 4 \pi^2 v_w g_\ast T } 
\int_0^\infty \mathrm{d}z \, \mu_{B_L} f_\mathrm{sph} \exp\Bigl( - \frac{45 \Gamma_\mathrm{sph} z}{  4 v_w} \Bigr), 
\end{equation}
where $g_\ast = 106.75$ is the effective degree of freedom for the entropy density at the electroweak phase transition. 
The symbol $\Gamma_\mathrm{sph}$ is the rate of the weak sphaleron which is evaluated as $\Gamma_\mathrm{sph} \simeq 1.0 \times 10^{-6} T$ by the lattice calculation~\cite{Moore:2000mx}.
The function $\mu_{B_L}$ is defined by using the solution of the transport equations as follows: 
\begin{equation}
\mu_{B_L} = \frac{ 1 }{ 2 } ( 1 + 4 K_{1,t})\mu_t + \frac{ 1 }{ 2 } (1 + 4 K_{1,b}) \mu_b
- 2 K_{1,t} \mu_{t^c}. 
\end{equation}
In addition, the function $f_\mathrm{sph}(z)$ corresponds to the effect of the suppression of the weak sphaleron rate for $z>0$ caused by the nonzero VEVs, and according to Ref.~\cite{Cline:2011mm}, it can be evaluated by
\begin{equation}
f_\mathrm{sph} = \mathrm{min}
\Bigl( 1, \ \frac{ 2.4 T }{ \Gamma_\mathrm{sph} } e^{-40 v(z) / T} \Bigr), 
\end{equation}
where $v(z) = \sqrt{ \hat{\varphi}_1^2(z) + \hat{\varphi}_2^2(z)  + \hat{\varphi}_3^2(z) }$.

\section{Numerical evaluation for the baryon asymmetry of the universe} 
\label{sec:evaluation}
 In this section, we show the results of our numerical evaluations for the phase transition and the baryon asymmetry in the model.  
For the parameters in the SM, the values at the scale of the mass of the $Z$ boson are used. They are shown in Table~\ref{table:SM_inputs}. 

\begin{table}[b]
\begin{center}
\begin{tabular}{| l l l l |} \hline
$m_u = 1.29 \times 10^{-3}$, \quad & $m_c = 0.619$, \quad & $m_t = 171.7$, \quad & $m_W^{} = 80.379$  \\
$m_d = 2.93 \times 10^{-3}$, \quad & $m_s = 0.055$, \quad & $m_b = 2.89$, \quad & $m_Z^{} = 91.1876$\\
$m_e = 4.87 \times 10^{-4}$, \quad & $m_\mu = 0.103$, \quad & $m_\tau = 1.746$. \quad & (in GeV) \\ \hline
$\alpha = 1/127.955$, & $\alpha_S^{} = 0.1179$.  & &  \\ \hline
\end{tabular}
\caption{The input parameters of the SM parameters~\cite{Kanemura:2020ibp}. The masses of fermions and gauge bosons are given in GeV. The coupling constants $\alpha$ and $\alpha_S^{}$ are the fine structure constant of QED and QCD, respectively. }
\label{table:SM_inputs}
\end{center}
\end{table}

For the numerical calculation of some quantities on the electroweak phase transition, for example, the nucleation temperature $T_n$, 
we use CosmoTransitions~\cite{Wainwright:2011kj}, which is a set of python modules for calculations about the effective potential. 
For simplicity, we consider only the single-step phase transition. 

\begin{figure}[h]
    \centering
    \includegraphics[width=150mm]{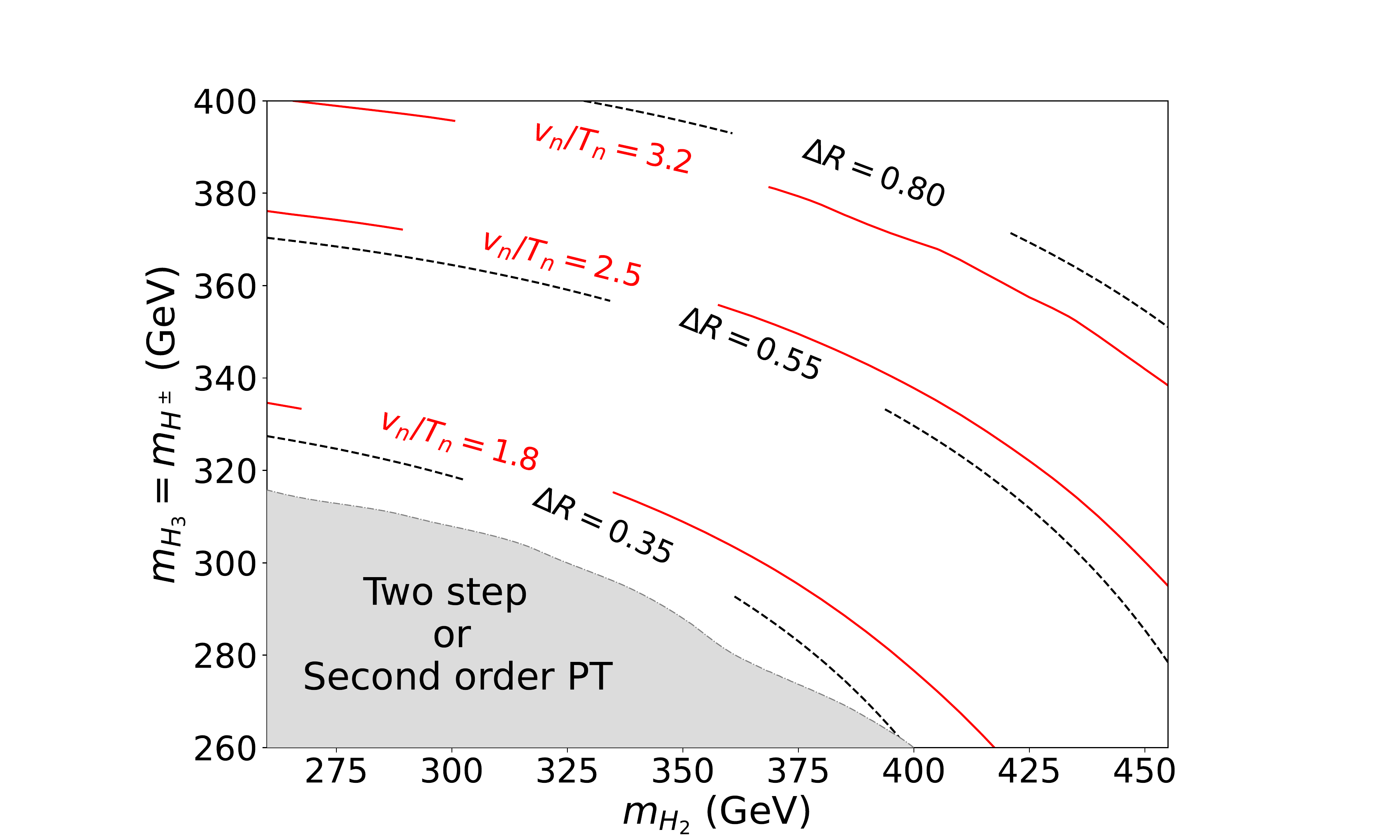}
    \caption{Contour plots of $v_n / T_n$ (red lines) and $\Delta R$ (black lines) on the plane of $m_{H_2}$ and $m_{H_3}$ ($= m_{H^\pm}$). The input parameters are shown in Eq.~(\ref{eq:Input1}). }
    \label{vnTn_DeltaR}
\end{figure}

For the electroweak baryogenesis, the condition $v_n/T_n \gtrsim 1$, where $v_n = v(-\infty)$ is the VEV at $T=T_n$, has to be satisfied 
in order to avoid washing out the generated baryon asymmetry~\cite{Kuzmin:1985mm}. 
In the model, this condition can be satisfied due to the non-decoupling effects of the additional scalar bosons $H^\pm$, $H_2$, and $H_3$. 
In such a case, it is known that a sizable deviation of the triple Higgs boson coupling from that in the SM ($\Delta R$ defined in Sec.~\ref{sec:effpot}) is expected~\cite{Kanemura:2004ch}. 
In Fig.~\ref{vnTn_DeltaR}, $v_n / T_n$ and $\Delta R$  are shown for various masses of additional scalar bosons. 
Input parameters for Fig.~\ref{vnTn_DeltaR} are as follows;
\begin{equation}
\label{eq:Input1}
\begin{array}{l}
M = 30~\mathrm{GeV}, \quad \lambda_2 = 0.1, \quad |\lambda_7| = 0.8, \quad 
\theta_7 = 0.9,  \\
|\zeta_e| = |\zeta_u| = |\zeta_d|  = 0.14, \quad 
\theta_u = \theta_d = 2.8, \quad 
\delta \equiv \theta_u - \theta_e = 0.05.  \\
\end{array}
\end{equation}
We have confirmed that these parameters satisfy all the theoretical and experimental constraints explained in Sec.~\ref{sec:constraints}. 
In the gray region in Fig.~\ref{vnTn_DeltaR}, 
the electroweak phase transition is the two-step one or the 2nd order one. 
We can see that in the region where the 1st order phase transition occurs, $v_n / T_n $ is enough large for the electroweak baryogenesis, and $\Delta R$ is expected to be about $30$--$80~\%$ in the mass region for the figure. 
Since the invariant mass parameter $M$ is fixed at $30~\mathrm{GeV}$, 
the non-decoupling effects of the additional scalars become larger in the heavier mass region, 
and $v_n/T_n$ and $\Delta R$ become larger too. 

\begin{figure}[h]
    \centering
    \includegraphics[width=150mm]{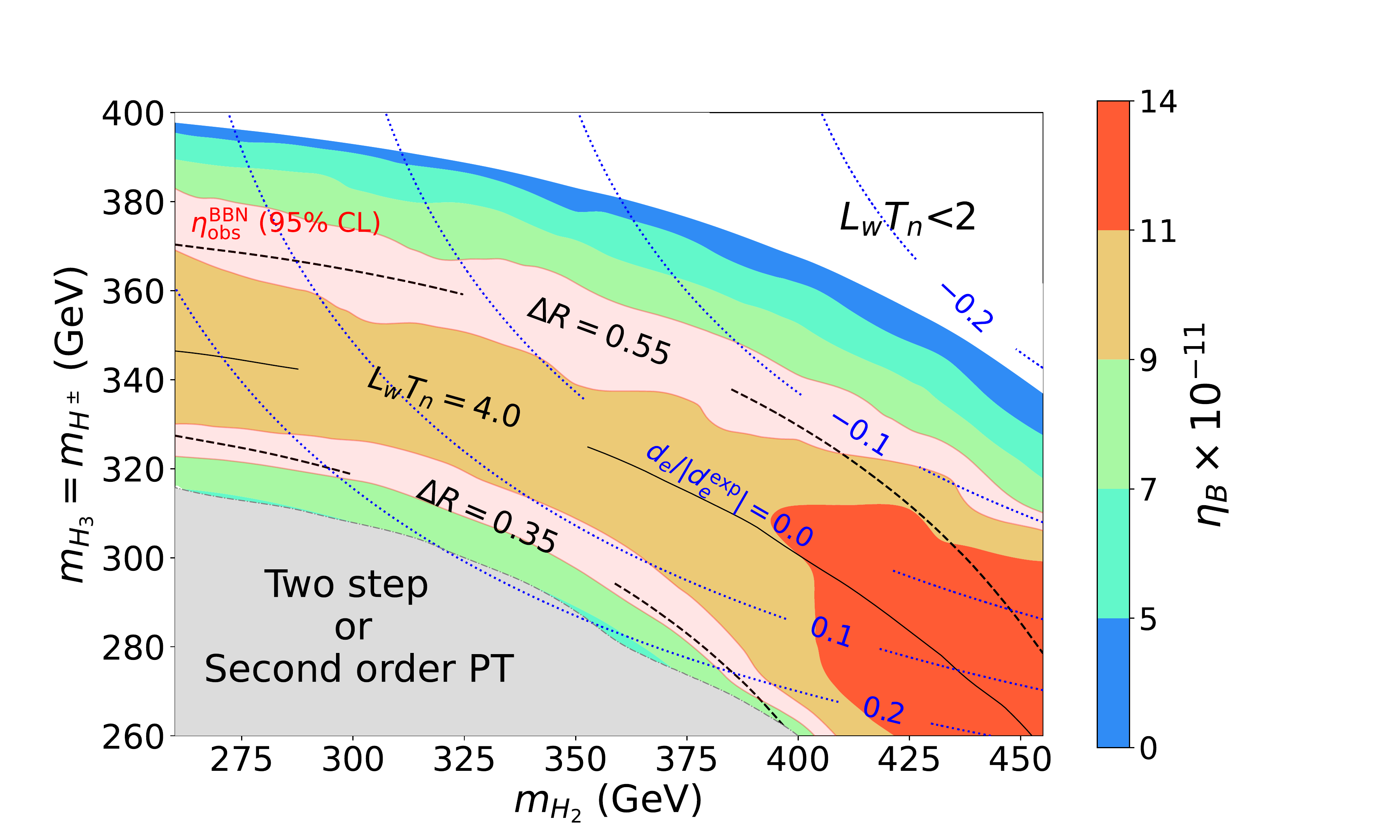}
    \caption{The generated baryon asymmetry is shown on the plane of $m_{H_2}$ and $m_{H_3}$ ($= m_{H^\pm}$). Input parameters are shown in Eq.~(\ref{eq:Input1}). On the point in the pink regions, the observed baryon asymmetry can be obtained. }
    \label{fig:Baryon_asymmetry}
\end{figure}

In Fig.~\ref{fig:Baryon_asymmetry}, generated baryon asymmetry via the electroweak phase transition is shown with the input parameters in Eq.~(\ref{eq:Input1}). 
The gray region in Fig.~\ref{fig:Baryon_asymmetry} is the same with that in Fig.~\ref{vnTn_DeltaR}. 
For the calculation of the baryon asymmetry $\eta_B$, 
the wall velocity $v_w$ is assumed to be just an input parameter and is fixed at $0.1$. 
The black real line in Fig.~\ref{fig:Baryon_asymmetry} is the contour of $L_w T_n = 4.0$, where $L_w$ is the width of the bubble wall and is determined by fitting the bounce solution of $v(z)$ with $\frac{v_n}{ 2} ( 1 - \tanh \frac{z}{ L_w } )$. 
In the region below this line, the generated baryon asymmetry increases as the masses of the additional scalar bosons increase because the phase transition is stronger as shown in Fig.~\ref{vnTn_DeltaR}. 
On the other hand, in the region above the line $L_w T_n = 4.0$, 
the baryon asymmetry decreases as the masses of the scalar bosons increase. 
This is because, as the phase transition is stronger, 
$L_w T_n$ is smaller, and it leads to the smaller baryon asymmetry. 
This behavior is discussed in Ref.~\cite{Cline:2021dkf}. 
In addition, for small $L_w T_n$, the WKB approach is not an appropriate approximation~\cite{Joyce:1994fu, Joyce:1994zn, Cline:2000nw}. 
In the two pink regions, the observed baryon asymmetry can be explained with better than $95~\%$ C.L. 
In these regions, $\Delta R$ is expected to be $35$--$55~\%$, 
and it would be tested at the HL-LHC~\cite{Cepeda:2019klc} and the future upgraded ILC~\cite{Fujii:2015jha, Bambade:2019fyw} and CLIC~\cite{CLICdp:2018cto}. 
In the orange and red regions sandwiched between the pink regions, the baryon number is overproduced. 

The blue dotted lines in Fig.~\ref{fig:Baryon_asymmetry} are contour plots of the eEDM for $d_e / |d_e^{\sf exp}| = 0$, $\pm 0.1$, and $\pm 0.2$, where $|d_e^{\sf exp}|$ ($=1.0 \times 10^{-29}~\mathrm{e\, cm}$) is the current upper limit on $|d_e|$~\cite{ACME:2018yjb}.
On the line of $|d_e|/|d_e^{\sf exp}| = 0.0$, two kinds of Barr-Zee type diagrams in Figs.~\ref{fig:BZf}~and~\ref{fig:BZs} cancel each other. 
The contours for the current upper limit $d_e / |d_e^{\sf exp}| = \pm 1.0$ are located outside of Fig.~\ref{fig:Baryon_asymmetry}. 
Therefore, in all the pink regions in Fig.~\ref{fig:Baryon_asymmetry}, 
the observed baryon asymmetry can be reproduced under the current eEDM constraint. 
At future eEDM experiments, the upper limit is expected to be improved by an order of magnitude~\cite{ACME:2018yjb}. 
Then, the region above the line of $d_e / |d_e^{\sf exp}| = -0.1$ and that below the line of $d_e / |d_e^{\sf exp}| = 0.1$ will be excluded. 
Almost all of the lower pink region can be thus tested by near-future improvement of the eEDM experiment. 
The nEDM is about four orders smaller than the current upper limit in the mass region shown in Fig.~\ref{fig:Baryon_asymmetry} because $\theta_u = \theta_d$. 
If $\theta_d$ is different from $\theta_u$, 
the nEDM increases, however even in this case, 
the nEDM is about one order smaller than the upper limit. 

\begin{figure}[h]
\begin{center}
\includegraphics[width=70mm]{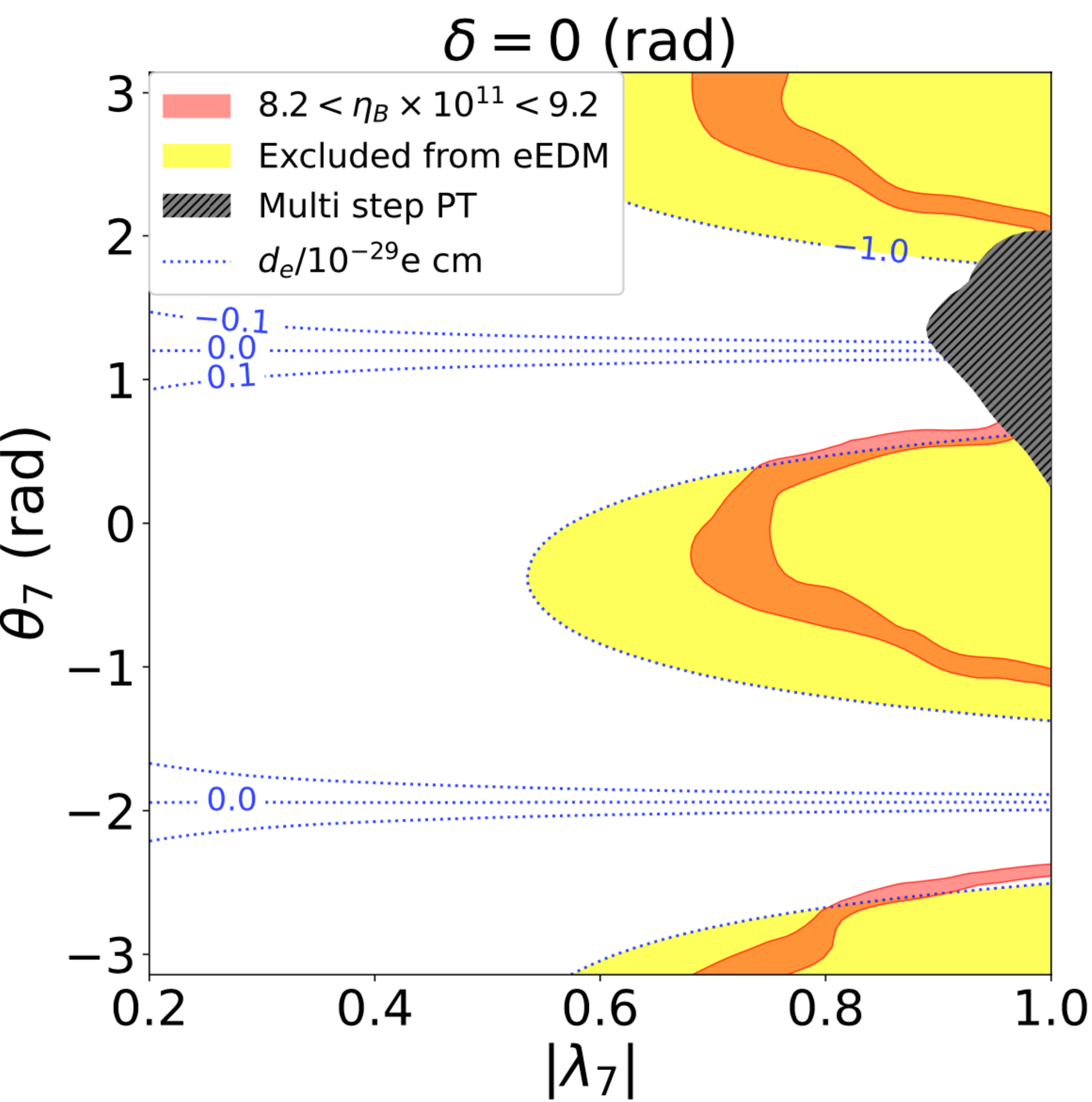}
\end{center}
\begin{minipage}[t]{0.45\textwidth}
\centering
\includegraphics[width=70mm]{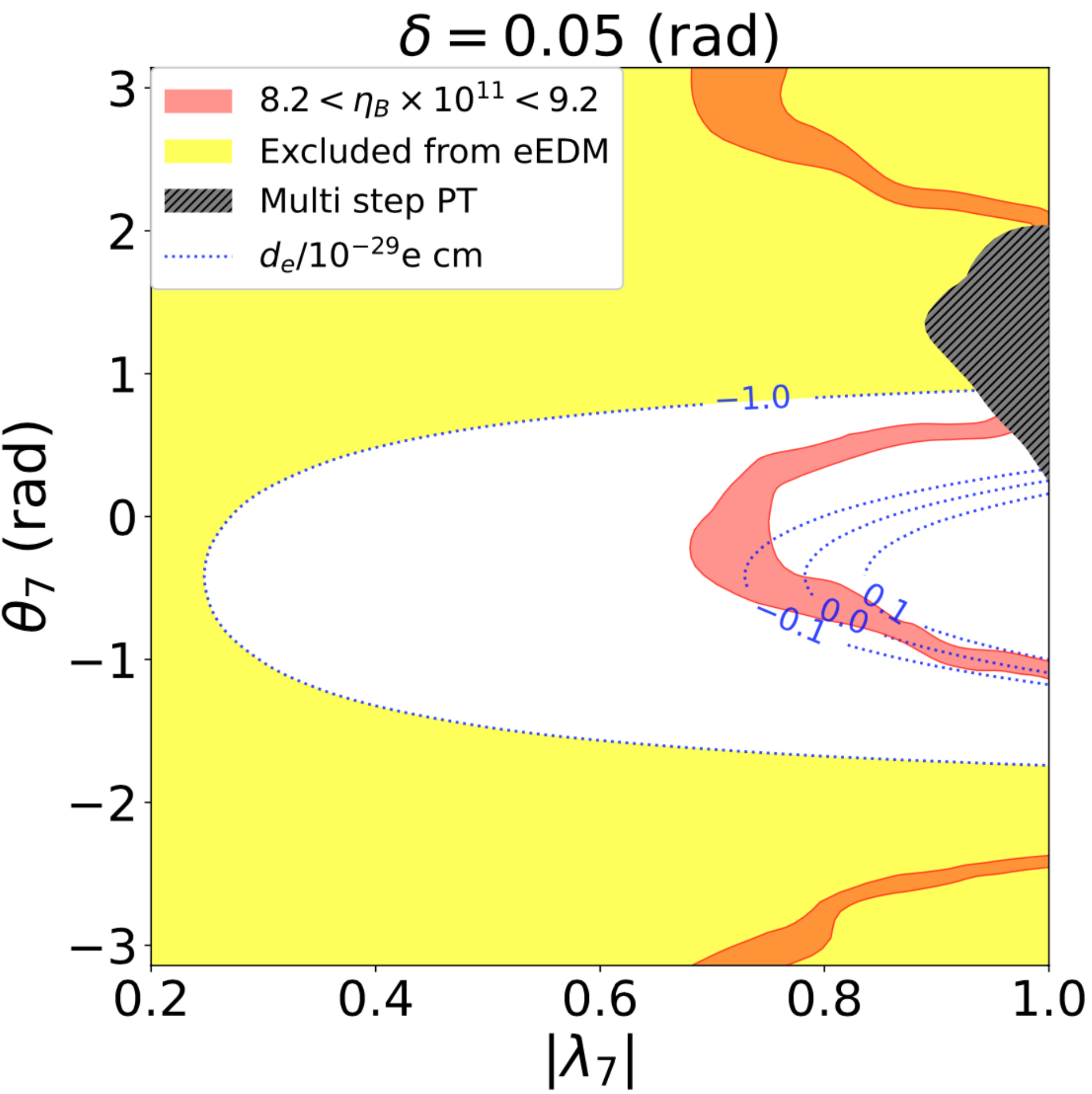}
\end{minipage}
\hspace{10pt}
\begin{minipage}[t]{0.45\textwidth}
\centering
\includegraphics[width=70mm]{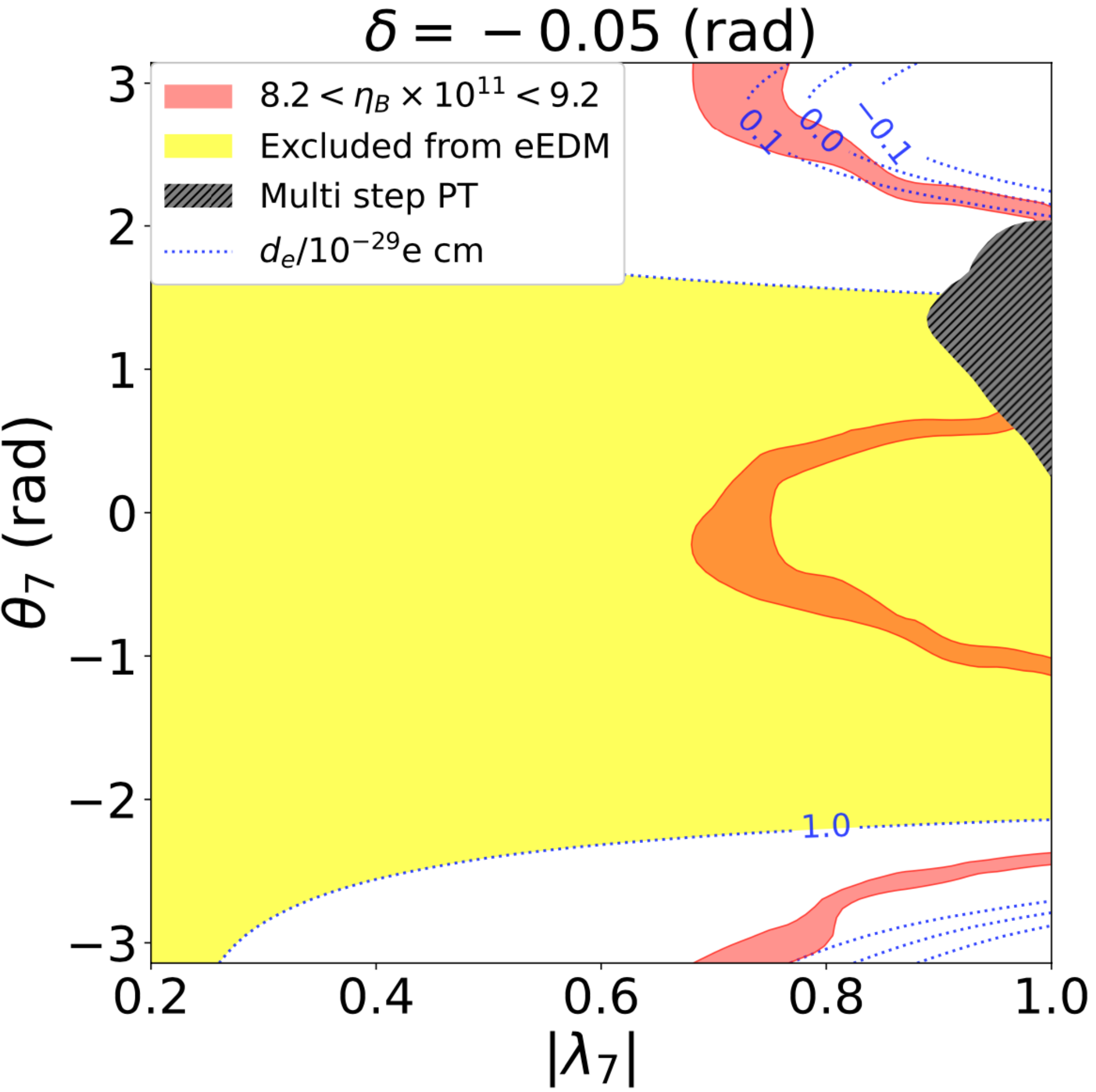}
\end{minipage}
\caption{Dependence on $\lambda_7$ of the eEDM and that of the baryon asymmetry for $\delta = 0$ (upper), $\delta = 0.05$ (lower left) and $\delta = -0.05$ (lower right). Input parameters are shown in Eq.~(\ref{eq:input2}). On the points in the pink regions, the observed baryon asymmetry can be obtained.}    
\label{fig:lambda7_dependence}
\end{figure}

In Fig.~\ref{fig:lambda7_dependence}, dependence on $\lambda_7$ of the eEDM and that of the baryon asymmetry are shown. 
The upper, lower left, and lower right figures are for $\delta$ ($\equiv \theta_u - \theta_e$) $= 0$, $\delta =0.05$, 
and $\delta = -0.05$, respectively. 
Input parameters for Fig.~\ref{fig:lambda7_dependence} are as follows;
\begin{equation}
\label{eq:input2}
\begin{array}{l}
m_{H_2} = m_{H_3} = m_{H^\pm} = 330~\mathrm{GeV}, \quad 
M = 30~\mathrm{GeV},  \quad
\lambda_2 = 0.1, \\
|\zeta_u| = |\zeta_d| = |\zeta_e| = 0.15, \quad 
\theta_u = \theta_d = 1.2.
\end{array} 
\end{equation}
In the black regions, the electroweak phase transition occurs in multiple steps, 
and we do not consider these points for simplicity. 
A typical value of $v_n / T_n$ on points in Fig.~\ref{fig:lambda7_dependence} 
is about $2.0$, and in the pink regions, the observed baryon asymmetry can be obtained. 
The pink regions in three figures are almost the same because the value of $\zeta_u$ is fixed and the generated baryon asymmetry is almost independent of $\zeta_e$. 

The blue lines in Fig.~\ref{fig:lambda7_dependence} are contour plot for the eEDM $d_e /|d_e^{\exp}| = 0$, $1.0$ and $-1.0$. 
In the yellow regions, $|d_e|$ is larger than $1.0 \times 10^{-29}~\mathrm{e\ cm}$, and it is excluded by the current eEDM experiment. 
In the upper figure ($\delta =0$), the contribution from the Barr-Zee type diagram including the top quark loop vanishes because $\sin \delta = 0$. 
Thus, the eEDM at two-loop level is $0$ on the lines of $\theta_7 - \theta_e = 0$ and $\theta_7 - \theta_e = \pi$, 
where the contribution from Barr-Zee type diagrams including scalar loops also vanishes. 
In this case, almost all pink regions are excluded by the current upper limit, and the remaining small area will also be excluded by one order improvement of the upper limit at future experiments of the eEDM. 

In the lower left figure ($\delta =0.05$), the Barr-Zee type diagram including the top quark loop gives a large contribution to the eEDM, and the excluded region by the current eEDM data is drastically changed. 
Even in this case, there are some points where the two kinds of Barr-Zee type diagrams are completely canceled by each other in the region $|\lambda_7| \gtrsim 0.8$.  
The pink regions for $\theta_7 \gtrsim 2$ and $\theta_7 \lesssim -2$
are excluded by the current eEDM experiments, while those for $-1 \lesssim \theta_7 \lesssim 1$ is allowed. 
By improvement of one order of magnitude in future EDM observations, 
the wide part of the current allowed region can be explored. 

The lower right figure is for $\delta = -0.05$. 
In this case, the sign of the Barr-Zee type diagram including the top quark loop is opposite from that for $\delta = 0.05$.
On the other hand, the absolute value is the same with that for $\delta = 0.05$. Therefore, the allowed region (white region) is shifted by $\pi$ in the direction of $\theta_7$ from that in the lower left figure ($\delta = 0.05$), since the Barr-Zee type diagrams including the scalar loop are proportional to $\sin (\theta_7 - \theta_e)$. 
The pink region for $-1 \lesssim \theta_7 \lesssim 1$ is excluded by the current eEDM experiments while those for $\theta_7 \gtrsim 2$ and $\theta_7 \lesssim -2$ still survive. 
By improvement of one order of magnitude in future EDM observations, 
most of the pink region for $\theta_7 \lesssim -2$ and the wide parts of that for $\theta_7 \gtrsim 2$ can be explored.

\section{Discussions}
\label{sec:discussion} 

In this section, we give some comments on the evaluation of the generated baryon asymmetry and the testability of the model in future experiments.  

In the numerical evaluations in Sec.~\ref{sec:evaluation}, 
$\zeta$-parameters are assumed to be $|\zeta_u| = |\zeta_d| = |\zeta_e|$ for simplicity. 
In this case, the top quarks play an important role in generating the BAU while the other quarks and leptons are negligible, because the Yukawa interactions respect the mass hierarchy of the SM fermions. 
This situation can be changed by relaxing the assumption $|\zeta_u| = |\zeta_d| = |\zeta_e|$. 
For example, in the case of $|\zeta_u| \ll |\zeta_e|$, tau leptons can play an essential role for the baryogenesis~\cite{Chung:2009cb, DeVries:2018aul, Xie:2020wzn}.
On the other hand, bottom quarks can also provide the main source of the CP-violation~\cite{Modak_Senaha}. 
In our model with aligned THDM, these scenarios may also be possible, 
although it is out of the scope of this paper. 

Furthermore, it can be considered to extend the model so that it includes flavor non-diagonal Yukawa interactions between $\Phi_2$ and the SM fermions. 
In such a case, the flavor non-diagonal couplings are severely restricted by flavor experiments. 
Nevertheless, it is known that there are some parameter regions where some of the non-diagonal interactions can play an important role for EWBG, for example, $t$-$c$ mixing~\cite{Fuyuto_Senaha} or $\tau$-$\mu$ mixing~\cite{Chiang:2016vgf, Guo:2016ixx}. 

In deriving the transport equation in our model, quadratic and higher terms of $v_w$ are neglected for simplicity. 
Effects of the higher-order terms are investigated in Ref.~\cite{Cline:2020jre}.  
In the case that $v_w = 0.1$, 
it is expected to decrease the generated baryon asymmetry by about $10$--$20~\%$. 
Therefore, by taking into account the higher-order terms, the regions of overproduction of the BAU in Figs.~\ref{fig:Baryon_asymmetry} and~\ref{fig:lambda7_dependence} would be suitable for successful EWBG. 

In our analysis, the WKB approximation method has been used to evaluate the BAU generated via the electroweak phase transition. 
There is however another method, which is so-called the VEV Insertion Approximation (VIA)~\cite{Riotto:1995hh, Riotto:1997vy}. 
Analyses by using the WKB and the VIA methods are compared in Refs.~\cite{Cline:2020jre, Cline:2021dkf, Basler:2021kgq}. 
The generated BAU in the VIA method is orders of magnitude larger than that in the WKB approximation. 
Thus, in the case if we use the VIA, the regions of the correct value of the BAU would be changed from the region for the successful baryogenesis in Figs.~\ref{fig:Baryon_asymmetry} and~\ref{fig:lambda7_dependence}.  
Even in this case, it would be able to obtain the observed BAU with satisfying the constraints in Sec.~\ref{sec:constraints} by taking $|\zeta_u|$ and $|\lambda_7|$ to be smaller than our benchmark scenario with keeping the destructive interference for the eEDM. 

Next, we discuss how to test our benchmark scenario in the aligned THDM by various future experiments. 
As shown in Fig.~\ref{fig:Baryon_asymmetry}, the EWBG can be realized by introducing the additional scalar bosons, whose masses are about $300$-$400~\mathrm{GeV}$. 
At future hadron colliders such as the HL-LHC, 
these additional Higgs bosons could be detected by searches for processes such as $A \to \tau \tau$ and $H^\pm \to tb$~\cite{Kanemura:2011kx, Kanemura:2014dea, Arbey:2017gmh, Arhrib:2018ewj, Aiko:2020ksl}.
The testability strongly depends on the hierarchy among $\zeta$-parameters. 
Furthermore, in this paper, we have assumed that the neutral scalar bosons do not mix with taking $\lambda_6=0$. 
If this alignment is slightly broken, 
decay branching ratios of the additional scalar bosons can drastically change due to the appearance of Higgs-to-Higgs decay processes, 
and the testability of the model at the HL-LHC can be much enhanced~\cite{Aiko:2020ksl}. 

The additional Higgs bosons can also be tested at future flavor experiments such as Belle-II~\cite{Belle-II:2018jsg} and LHCb~\cite{LHCb:2012myk} by observing the processes like $B \to X_s \gamma$, $B_s \to \mu^+ \mu^-$, and the $B^0$-$\overline{B^0}$ mixing. 
In addition, by searching for the asymmetry between the CP-violation in $B^- \to X_s^- \gamma$ and that in $B^0 \to X_s^0 \gamma$,  
the CP-violating parameters $\zeta_u$ and $\zeta_d$ would be tested~\cite{Benzke:2010tq, Belle:2018iff, Modak_Senaha}. 

The CP-violating phases in the Yukawa interaction can be tested by the eEDM and the nEDM as mentioned in Sec.~\ref{sec:constraints}. 
The future eEDM experiment is expected to improve the upper limit on $|d_e|$ by one order~\cite{ACME:2018yjb}. 
As shown in Figs.~\ref{fig:Baryon_asymmetry} and~\ref{fig:lambda7_dependence}, 
by this improvement, most of the regions in the figures can be tested. 
The upper limit on the nEDM is also expected to be improved by one order of magnitude at future experiments~\cite{Martin:2020lbx}. 
As mentioned in Sec.~\ref{sec:evaluation}, 
in the regions for the successful electroweak baryogenesis in Figs.~\ref{fig:Baryon_asymmetry} and~\ref{fig:lambda7_dependence}, 
the maximal value of the nEDM is one order smaller than the current upper limit. 
Therefore, the parameter regions of our model would be further tested by future nEDM experiments. 

In addition, if the additional neutral Higgs bosons are detected at the HL-LHC or at future lepton colliders, 
the CP-violating coupling $\zeta_e$ would be tested by observing the azimuthal angle distribution of the hadronic decay of the tau leptons from the decays of $H_2$ and $H_3$ at future $e^+$-$e^-$ colliders such as the ILC~\cite{Jeans:2018anq, Kanemura:2021atq}.
This would be useful to test our model especially in the case that $|\zeta_e| \gg |\zeta_u|$, $|\zeta_d|$ and $\theta_e =\mathcal{O}(1)$.

The strongly 1st order phase transition in our model can be tested by measuring the deviation in the triple Higgs boson coupling via the di-Higgs production at future high-energy colliders~\cite{Pairprod_had1, Pairprod_had2, Goncalves:2018qas, Tian:2013qmi, Kurata:2013, Fujii:2015jha, Asakawa:2010xj}.  
At the HL-LHC and the upgraded ILC with $\sqrt{s} = 500~\mathrm{GeV}$ ($1~\mathrm{TeV}$), the triple Higgs boson coupling is expected to be measured by the accuracies of $50~\%$~\cite{Cepeda:2019klc} and $27~\%$ ($10~\%$)~\cite{Fujii:2015jha, Bambade:2019fyw}, respectively.
As shown in Figs.~\ref{vnTn_DeltaR} and~\ref{fig:Baryon_asymmetry}, 
the deviation in the triple Higgs boson coupling, $\Delta R$, is predicted to be $35$ - $55~\%$ for successful EWBG. 
These deviations would be tested at the HL-LHC and the future upgraded ILC.

Furthermore, GWs can also be produced from the 1st order electroweak phase transition in extended Higgs sectors~\cite{Kakizaki:2015wua}. 
They would be detectable at future space-based GW detectors such as LISA~\cite{LISA:2017pwj}, DECIGO~\cite{Seto:2001qf} and BBO~\cite{Corbin:2005ny}. 
Detailed analyses of this possibility will be performed in our model elsewhere~\cite{Enomoto_future}.

\section{Conclusions}
\label{sec:conclusion}

In this paper, we have evaluated the BAU in the charge transport scenario of top quarks in the CP-violating THDM, in which Yukawa interactions are aligned to avoid dangerous FCNCs, and coupling constants of the lightest Higgs boson with the mass $125~\mathrm{GeV}$ coincide with those in the standard model at tree level to satisfy the current LHC data. 
In this model, the severe constraint from the electric dipole moment of electrons, which are normally difficult to be satisfied, can be avoided by destructive interferences between CP-violating phases in Yukawa interactions and scalar couplings in the Higgs potential. 
We have proposed viable benchmark scenarios for EWBG in this model under the current available data and basic theoretical bounds. 
We have found that the observed baryon number can be reproduced in this model, where masses of additional Higgs bosons are typically $300$-$400~\mathrm{GeV}$. 
These additional Higgs bosons can be directly discovered at future collider experiments. 
The effects of CP-violation in our benchmark scenario can also be tested at future EDM experiments, flavor experiments and $e^+$-$e^-$ collider experiments. 
Furthermore, the triple Higgs boson coupling is predicted to be $35$-$55~\%$ larger than the standard model value, which can also be tested at future hadron and lepton colliders.

\section*{Acknowledgement}
The work of K.~E. was supported in part by JSPS KAKENHI Grant No.~JP21J11444. The work of S.~K. was supported in part by the Grant-in-Aid on Innovative Areas, the Ministry of Education, Culture, Sports, Science and Technology, No.~16H06492 and JSPS KAKENHI Grant No.~20H00160.


\begin{thebibliography}{99}


\bibitem{ParticleDataGroup:2020ssz}
P.~A.~Zyla \textit{et al.} [Particle Data Group],
``Review of Particle Physics,''
PTEP \textbf{2020} (2020) no.8, 083C01. 

\bibitem{Sakharov:1967dj}
A.~D.~Sakharov,
``Violation of CP Invariance, C asymmetry, and baryon asymmetry of the universe,''
Pisma Zh. Eksp. Teor. Fiz. \textbf{5} (1967), 32-35. 

\bibitem{Huet:1994jb}
P.~Huet and E.~Sather,
``Electroweak baryogenesis and standard model CP violation,''
Phys. Rev. D \textbf{51} (1995), 379-394
[arXiv:hep-ph/9404302 [hep-ph]].

\bibitem{EWPT_SM}
K.~Kajantie, M.~Laine, K.~Rummukainen and M.~E.~Shaposhnikov,
``Is there a~ hot electroweak phase transition at $m_H \gtrsim m_W$?,''
Phys. Rev. Lett. \textbf{77} (1996), 2887-2890
[arXiv:hep-ph/9605288 [hep-ph]]; 
%
M.~D'Onofrio and K.~Rummukainen,
``Standard model cross-over on the lattice,''
Phys. Rev. D \textbf{93} (2016) no.2, 025003
[arXiv:1508.07161 [hep-ph]].

\bibitem{GUT_Baryogenesis}
M.~Yoshimura,
``Unified Gauge Theories and the Baryon Number of the Universe,''
Phys. Rev. Lett. \textbf{41} (1978), 281-284
[erratum: Phys. Rev. Lett. \textbf{42} (1979), 746]; 
%
S.~Weinberg,
``Cosmological Production of Baryons,''
Phys. Rev. Lett. \textbf{42} (1979), 850-853. 

\bibitem{Affleck:1984fy}
I.~Affleck and M.~Dine,
``A New Mechanism for Baryogenesis,''
Nucl. Phys. B \textbf{249} (1985), 361-380. 

\bibitem{Kuzmin:1985mm}
V.~A.~Kuzmin, V.~A.~Rubakov and M.~E.~Shaposhnikov,
``On the Anomalous Electroweak Baryon Number Nonconservation in the Early Universe,''
Phys. Lett. B \textbf{155} (1985), 36. 

\bibitem{Fukugita:1986hr}
M.~Fukugita and T.~Yanagida,
``Baryogenesis Without Grand Unification,''
Phys. Lett. B \textbf{174} (1986), 45-47. 

\bibitem{Klinkhamer:1984di}
F.~R.~Klinkhamer and N.~S.~Manton,
``A Saddle Point Solution in the Weinberg-Salam Theory,''
Phys. Rev. D \textbf{30} (1984), 2212

\bibitem{Turok:1990in}
N.~Turok and J.~Zadrozny,
``Dynamical generation of baryons at the electroweak transition,''
Phys. Rev. Lett. \textbf{65} (1990), 2331-2334

\bibitem{Turok:1990zg}
N.~Turok and J.~Zadrozny,
``Electroweak baryogenesis in the two doublet model,''
Nucl. Phys. B \textbf{358} (1991), 471-493

\bibitem{Cline:1995dg}
J.~M.~Cline, K.~Kainulainen and A.~P.~Vischer,
``Dynamics of two Higgs doublet CP violation and baryogenesis at the electroweak phase transition,''
Phys. Rev. D \textbf{54} (1996), 2451-2472
[arXiv:hep-ph/9506284 [hep-ph]].

\bibitem{Fromme:2006cm}
L.~Fromme, S.~J.~Huber and M.~Seniuch,
``Baryogenesis in the two-Higgs doublet model,''
JHEP \textbf{11} (2006), 038
[arXiv:hep-ph/0605242 [hep-ph]].

\bibitem{Cline:2011mm}
J.~M.~Cline, K.~Kainulainen and M.~Trott,
``Electroweak Baryogenesis in Two Higgs Doublet Models and B meson anomalies,''
JHEP \textbf{11} (2011), 089
[arXiv:1107.3559 [hep-ph]].

\bibitem{Tulin:2011wi}
S.~Tulin and P.~Winslow,
``Anomalous $B$ meson mixing and baryogenesis,''
Phys. Rev. D \textbf{84} (2011), 034013
[arXiv:1105.2848 [hep-ph]].

\bibitem{Liu:2011jh}
T.~Liu, M.~J.~Ramsey-Musolf and J.~Shu,
``Electroweak Beautygenesis: From $b \to s$ CP-violation to the Cosmic Baryon Asymmetry,''
Phys. Rev. Lett. \textbf{108} (2012), 221301
[arXiv:1109.4145 [hep-ph]].

\bibitem{Ahmadvand:2013sna}
M.~Ahmadvand,
``Baryogenesis within the two-Higgs-doublet model in the Electroweak scale,''
Int. J. Mod. Phys. A \textbf{29} (2014) no.20, 1450090
[arXiv:1308.3767 [hep-ph]].

\bibitem{Chiang:2016vgf}
C.~W.~Chiang, K.~Fuyuto and E.~Senaha,
``Electroweak Baryogenesis with Lepton Flavor Violation,''
Phys. Lett. B \textbf{762} (2016), 315-320
[arXiv:1607.07316 [hep-ph]].

\bibitem{Guo:2016ixx}
H.~K.~Guo, Y.~Y.~Li, T.~Liu, M.~Ramsey-Musolf and J.~Shu,
``Lepton-Flavored Electroweak Baryogenesis,''
Phys. Rev. D \textbf{96} (2017) no.11, 115034
[arXiv:1609.09849 [hep-ph]].

\bibitem{Fuyuto_Senaha}
K.~Fuyuto, W.~S.~Hou and E.~Senaha,
``Electroweak baryogenesis driven by extra top Yukawa couplings,''
Phys. Lett. B \textbf{776} (2018), 402-406
[arXiv:1705.05034 [hep-ph]]; 
``Cancellation mechanism for the electron electric dipole moment connected with the baryon asymmetry of the Universe,''
Phys. Rev. D \textbf{101} (2020) no.1, 011901
[arXiv:1910.12404 [hep-ph]].

\bibitem{Modak_Senaha}
T.~Modak and E.~Senaha,
``Electroweak baryogenesis via bottom transport,''
Phys. Rev. D \textbf{99} (2019) no.11, 115022
[arXiv:1811.08088 [hep-ph]]; 
``Probing Electroweak Baryogenesis induced by extra bottom Yukawa coupling via EDMs and collider signatures,''
JHEP \textbf{11} (2020), 025
[arXiv:2005.09928 [hep-ph]]; 
``Electroweak baryogenesis via bottom transport: Complementarity between LHC and future lepton collider probes,''
Phys. Lett. B \textbf{822} (2021), 136695
[arXiv:2107.12789 [hep-ph]].

\bibitem{Basler:2021kgq}
P.~Basler, M.~M\"uhlleitner and J.~M\"uller,
``Electroweak Baryogenesis in the CP-Violating Two-Higgs Doublet Model,''
[arXiv:2108.03580 [hep-ph]].

\bibitem{Huet:1995sh}
P.~Huet and A.~E.~Nelson,
``Electroweak baryogenesis in supersymmetric models,''
Phys. Rev. D \textbf{53} (1996), 4578-4597
[arXiv:hep-ph/9506477 [hep-ph]].

\bibitem{Cline:2000nw}
J.~M.~Cline, M.~Joyce and K.~Kainulainen,
``Supersymmetric electroweak baryogenesis,''
JHEP \textbf{07} (2000), 018
[arXiv:hep-ph/0006119 [hep-ph]]; 

\bibitem{Cirigliano:2006dg}
V.~Cirigliano, S.~Profumo and M.~J.~Ramsey-Musolf,
``Baryogenesis, Electric Dipole Moments and Dark Matter in the MSSM,''
JHEP \textbf{07} (2006), 002
[arXiv:hep-ph/0603246 [hep-ph]].

\bibitem{Espinosa:2011eu}
J.~R.~Espinosa, B.~Gripaios, T.~Konstandin and F.~Riva,
``Electroweak Baryogenesis in Non-minimal Composite Higgs Models,''
JCAP \textbf{01} (2012), 012
[arXiv:1110.2876 [hep-ph]].

\bibitem{Cline:2012hg}
J.~M.~Cline and K.~Kainulainen,
``Electroweak baryogenesis and dark matter from a singlet Higgs,''
JCAP \textbf{01} (2013), 012
[arXiv:1210.4196 [hep-ph]].

\bibitem{Bodeker:2004ws}
D.~Bodeker, L.~Fromme, S.~J.~Huber and M.~Seniuch,
``The Baryon asymmetry in the standard model with a low cut-off,''
JHEP \textbf{02} (2005), 026
[arXiv:hep-ph/0412366 [hep-ph]].

\bibitem{Fromme:2006wx}
L.~Fromme and S.~J.~Huber,
``Top transport in electroweak baryogenesis,''
JHEP \textbf{03} (2007), 049
[arXiv:hep-ph/0604159 [hep-ph]].

\bibitem{Cline:2020jre}
J.~M.~Cline and K.~Kainulainen,
``Electroweak baryogenesis at high bubble wall velocities,''
Phys. Rev. D \textbf{101} (2020) no.6, 063525
[arXiv:2001.00568 [hep-ph]].

\bibitem{Pilaftsis:2002fe}
A.~Pilaftsis,
``Higgs mediated electric dipole moments in the MSSM: An application to baryogenesis and Higgs searches,''
Nucl. Phys. B \textbf{644} (2002), 263-289
[arXiv:hep-ph/0207277 [hep-ph]].

\bibitem{Huber:2006ri}
S.~J.~Huber, M.~Pospelov and A.~Ritz,
``Electric dipole moment constraints on minimal electroweak baryogenesis,''
Phys. Rev. D \textbf{75} (2007), 036006
[arXiv:hep-ph/0610003 [hep-ph]].

\bibitem{Ipek:2013iba}
S.~Ipek,
``Perturbative analysis of the electron electric dipole moment and CP violation in two-Higgs-doublet models,''
Phys. Rev. D \textbf{89} (2014) no.7, 073012
[arXiv:1310.6790 [hep-ph]].

\bibitem{Bian:2014zka}
L.~Bian, T.~Liu and J.~Shu,
``Cancellations Between Two-Loop Contributions to the Electron Electric Dipole Moment with a CP-Violating Higgs Sector,''
Phys. Rev. Lett. \textbf{115} (2015), 021801
[arXiv:1411.6695 [hep-ph]].

\bibitem{Kanemura:2020ibp}
S.~Kanemura, M.~Kubota and K.~Yagyu,
``Aligned CP-violating Higgs sector canceling the electric dipole moment'',
JHEP \textbf{08} (2020), 026
[arXiv:2004.03943 [hep-ph]].

\bibitem{Turok:1991uc}
N.~Turok and J.~Zadrozny,
``Phase transitions in the two doublet model,''
Nucl. Phys. B \textbf{369} (1992), 729-742

\bibitem{Anderson:1991zb}
G.~W.~Anderson and L.~J.~Hall,
``The Electroweak phase transition and baryogenesis,''
Phys. Rev. D \textbf{45} (1992), 2685-2698

\bibitem{Land:1992sm}
D.~Land and E.~D.~Carlson,
``Two stage phase transition in two Higgs models,''
Phys. Lett. B \textbf{292} (1992), 107-112
[arXiv:hep-ph/9208227 [hep-ph]].

\bibitem{Hammerschmitt:1994fn}
A.~Hammerschmitt, J.~Kripfganz and M.~G.~Schmidt,
``Baryon asymmetry from a two stage electroweak phase transition?,''
Z. Phys. C \textbf{64} (1994), 105-110
[arXiv:hep-ph/9404272 [hep-ph]].

\bibitem{Cline:1996mga}
J.~M.~Cline and P.~A.~Lemieux,
``Electroweak phase transition in two Higgs doublet models,''
Phys. Rev. D \textbf{55} (1997), 3873-3881
[arXiv:hep-ph/9609240 [hep-ph]].

\bibitem{Laine:2000rm}
M.~Laine and K.~Rummukainen,
``Two Higgs doublet dynamics at the electroweak phase transition: A Nonperturbative study,''
Nucl. Phys. B \textbf{597} (2001), 23-69
[arXiv:hep-lat/0009025 [hep-lat]].

\bibitem{Blinov:2015sna}
N.~Blinov, J.~Kozaczuk, D.~E.~Morrissey and C.~Tamarit,
``Electroweak Baryogenesis from Exotic Electroweak Symmetry Breaking,''
Phys. Rev. D \textbf{92} (2015) no.3, 035012
[arXiv:1504.05195 [hep-ph]].

\bibitem{Inoue:2015pza}
S.~Inoue, G.~Ovanesyan and M.~J.~Ramsey-Musolf,
``Two-Step Electroweak Baryogenesis,''
Phys. Rev. D \textbf{93} (2016), 015013
[arXiv:1508.05404 [hep-ph]].

\bibitem{Basler:2016obg}
P.~Basler, M.~Krause, M.~Muhlleitner, J.~Wittbrodt and A.~Wlotzka,
``Strong First Order Electroweak Phase Transition in the CP-Conserving 2HDM Revisited,''
JHEP \textbf{02} (2017), 121
[arXiv:1612.04086 [hep-ph]].

\bibitem{Andersen:2017ika}
J.~O.~Andersen, T.~Gorda, A.~Helset, L.~Niemi, T.~V.~I.~Tenkanen, A.~Tranberg, A.~Vuorinen and D.~J.~Weir,
``Nonperturbative Analysis of the Electroweak Phase Transition in the Two Higgs Doublet Model,''
Phys. Rev. Lett. \textbf{121} (2018) no.19, 191802
[arXiv:1711.09849 [hep-ph]].

\bibitem{Aoki:2021oez}
M.~Aoki, T.~Komatsu and H.~Shibuya,
``Possibility of multi-step electroweak phase transition in the two Higgs doublet models,''
[arXiv:2106.03439 [hep-ph]].

\bibitem{Kanemura:2004ch}
S.~Kanemura, Y.~Okada and E.~Senaha,
``Electroweak baryogenesis and quantum corrections to the triple Higgs boson coupling,''
Phys. Lett. B \textbf{606} (2005), 361-366
[arXiv:hep-ph/0411354 [hep-ph]].

\bibitem{Kanemura:2002vm}
S.~Kanemura, S.~Kiyoura, Y.~Okada, E.~Senaha and C.~P.~Yuan,
``New physics effect on the Higgs selfcoupling,''
Phys. Lett. B \textbf{558} (2003), 157-164
[arXiv:hep-ph/0211308 [hep-ph]].

\bibitem{Kanemura:2004mg}
S.~Kanemura, Y.~Okada, E.~Senaha and C.~P.~Yuan,
``Higgs coupling constants as a probe of new physics,''
Phys. Rev. D \textbf{70} (2004), 115002
[arXiv:hep-ph/0408364 [hep-ph]].

\bibitem{Braathen_Kanemura}
J.~Braathen and S.~Kanemura,
``On two-loop corrections to the Higgs trilinear coupling in models with extended scalar sectors,''
Phys. Lett. B \textbf{796} (2019), 38-46
[arXiv:1903.05417 [hep-ph]];
``Leading two-loop corrections to the Higgs boson self-couplings in models with extended scalar sectors,''
Eur. Phys. J. C \textbf{80} (2020) no.3, 227
[arXiv:1911.11507 [hep-ph]].

\bibitem{Pairprod_had1}
O.~J.~P.~Eboli, G.~C.~Marques, S.~F.~Novaes and A.~A.~Natale,
``TWIN HIGGS BOSON PRODUCTION,''
Phys. Lett. B \textbf{197} (1987), 269-272;
D.~A.~Dicus, C.~Kao and S.~S.~D.~Willenbrock,
``Higgs Boson Pair Production From Gluon Fusion,''
Phys. Lett. B \textbf{203} (1988), 457-461;
E.~W.~N.~Glover and J.~J.~van der Bij,
``HIGGS BOSON PAIR PRODUCTION VIA GLUON FUSION,''
Nucl. Phys. B \textbf{309} (1988), 282-294.

\bibitem{Pairprod_had2}
T.~Plehn, M.~Spira and P.~M.~Zerwas,
``Pair production of neutral Higgs particles in gluon-gluon collisions,''
Nucl. Phys. B \textbf{479} (1996), 46-64
[erratum: Nucl. Phys. B \textbf{531} (1998), 655-655]
[arXiv:hep-ph/9603205 [hep-ph]];
A.~Djouadi, W.~Kilian, M.~Muhlleitner and P.~M.~Zerwas,
``Production of neutral Higgs boson pairs at LHC,''
Eur. Phys. J. C \textbf{10} (1999), 45-49
[arXiv:hep-ph/9904287 [hep-ph]];
X.~Li and M.~B.~Voloshin,
``Remarks on double Higgs boson production by gluon fusion at threshold,''
Phys. Rev. D \textbf{89} (2014) no.1, 013012
[arXiv:1311.5156 [hep-ph]].

\bibitem{Goncalves:2018qas}
D.~Gon\c{c}alves, T.~Han, F.~Kling, T.~Plehn and M.~Takeuchi,
``Higgs boson pair production at future hadron colliders: From kinematics to dynamics,''
Phys. Rev. D \textbf{97} (2018) no.11, 113004
[arXiv:1802.04319 [hep-ph]].

\bibitem{Tian:2013qmi}
J.~Tian,
``Study of Higgs self-coupling at the ILC based on the full detector simulation at \ensuremath{\sqrt{s} = 500} GeV and \ensuremath{\sqrt{s} = 1} TeV,''

\bibitem{Kurata:2013}
M.~Kurata, T.~Tanabe, J.~Tian, K.~Fujii, T.~Suehara
``The Higgs Self Coupling Analysis Using The Events Containing $H\to WW^*$ Decay''.

\bibitem{Fujii:2015jha}
K.~Fujii, C.~Grojean, M.~E.~Peskin, T.~Barklow, Y.~Gao, S.~Kanemura, H.~D.~Kim, J.~List, M.~Nojiri and M.~Perelstein, \textit{et al.}
``Physics Case for the International Linear Collider,''
[arXiv:1506.05992 [hep-ex]].

\bibitem{Asakawa:2010xj}
E.~Asakawa, D.~Harada, S.~Kanemura, Y.~Okada and K.~Tsumura,
``Higgs boson pair production in new physics models at hadron, lepton, and photon colliders,''
Phys. Rev. D \textbf{82} (2010), 115002
[arXiv:1009.4670 [hep-ph]].


\bibitem{Apreda:2001us}
R.~Apreda, M.~Maggiore, A.~Nicolis and A.~Riotto,
``Gravitational waves from electroweak phase transitions,''
Nucl. Phys. B \textbf{631} (2002), 342-368
[arXiv:gr-qc/0107033 [gr-qc]].

\bibitem{Grojean:2006bp}
C.~Grojean and G.~Servant,
``Gravitational Waves from Phase Transitions at the Electroweak Scale and Beyond,''
Phys. Rev. D \textbf{75} (2007), 043507
[arXiv:hep-ph/0607107 [hep-ph]].

\bibitem{Huber:2007vva}
S.~J.~Huber and T.~Konstandin,
``Production of gravitational waves in the nMSSM,''
JCAP \textbf{05} (2008), 017
[arXiv:0709.2091 [hep-ph]].

\bibitem{Ashoorioon:2009nf}
A.~Ashoorioon and T.~Konstandin,
``Strong electroweak phase transitions without collider traces,''
JHEP \textbf{07} (2009), 086
[arXiv:0904.0353 [hep-ph]].

\bibitem{Kakizaki:2015wua}
M.~Kakizaki, S.~Kanemura and T.~Matsui,
``Gravitational waves as a probe of extended scalar sectors with the first order electroweak phase transition,''
Phys. Rev. D \textbf{92} (2015) no.11, 115007
[arXiv:1509.08394 [hep-ph]].

\bibitem{Huang:2015izx}
F.~P.~Huang, P.~H.~Gu, P.~F.~Yin, Z.~H.~Yu and X.~Zhang,
``Testing the electroweak phase transition and electroweak baryogenesis at the LHC and a circular electron-positron collider,''
Phys. Rev. D \textbf{93} (2016) no.10, 103515
[arXiv:1511.03969 [hep-ph]].

\bibitem{Hashino:2016rvx}
K.~Hashino, M.~Kakizaki, S.~Kanemura and T.~Matsui,
``Synergy between measurements of gravitational waves and the triple-Higgs coupling in probing the first-order electroweak phase transition,''
Phys. Rev. D \textbf{94} (2016) no.1, 015005
[arXiv:1604.02069 [hep-ph]].

\bibitem{Hashino:2016xoj}
K.~Hashino, M.~Kakizaki, S.~Kanemura, P.~Ko and T.~Matsui,
``Gravitational waves and Higgs boson couplings for exploring first order phase transition in the model with a singlet scalar field,''
Phys. Lett. B \textbf{766} (2017), 49-54
[arXiv:1609.00297 [hep-ph]].

\bibitem{Hashino:2018wee}
K.~Hashino, R.~Jinno, M.~Kakizaki, S.~Kanemura, T.~Takahashi and M.~Takimoto,
``Selecting models of first-order phase transitions using the synergy between collider and gravitational-wave experiments,''
Phys. Rev. D \textbf{99} (2019) no.7, 075011
[arXiv:1809.04994 [hep-ph]].

\bibitem{Pich:2009sp}
A.~Pich and P.~Tuzon,
``Yukawa Alignment in the Two-Higgs-Doublet Model,''
Phys. Rev. D \textbf{80} (2009), 091702
[arXiv:0908.1554 [hep-ph]].

\bibitem{CMS:2018uag}
A.~M.~Sirunyan \textit{et al.} [CMS],
``Combined measurements of Higgs boson couplings in proton\textendash{}proton collisions at $\sqrt{s}=13\,\text {Te}\text {V} $,''
Eur. Phys. J. C \textbf{79} (2019) no.5, 421
[arXiv:1809.10733 [hep-ex]].

\bibitem{ATLAS:2019nkf}
G.~Aad \textit{et al.} [ATLAS],
``Combined measurements of Higgs boson production and decay using up to $80$ fb$^{-1}$ of proton-proton collision data at $\sqrt{s}=$ 13 TeV collected with the ATLAS experiment,''
Phys. Rev. D \textbf{101} (2020) no.1, 012002
[arXiv:1909.02845 [hep-ex]].

\bibitem{Aiko:2020ksl}
M.~Aiko, S.~Kanemura, M.~Kikuchi, K.~Mawatari, K.~Sakurai and K.~Yagyu,
``Probing extended Higgs sectors by the synergy between direct searches at the LHC and precision tests at future lepton colliders,''
Nucl. Phys. B \textbf{966} (2021), 115375
[arXiv:2010.15057 [hep-ph]].

\bibitem{Kanemura:inpreperation}
S.~Kanemura, M.~Takeuchi, K.~Yagyu, in preperation. 

\bibitem{Kanemura:2021atq}
S.~Kanemura, M.~Kubota and K.~Yagyu,
``Testing aligned CP-violating Higgs sector at future lepton colliders,''
JHEP \textbf{04} (2021), 144
[arXiv:2101.03702 [hep-ph]].

\bibitem{ACME:2018yjb}
V.~Andreev \textit{et al.} [ACME],
``Improved limit on the electric dipole moment of the electron,''
Nature \textbf{562} (2018) no.7727, 355-360.

\bibitem{Joyce:1994fu}
M.~Joyce, T.~Prokopec and N.~Turok,
``Electroweak baryogenesis from a classical force,''
Phys. Rev. Lett. \textbf{75} (1995), 1695-1698
[erratum: Phys. Rev. Lett. \textbf{75} (1995), 3375]
[arXiv:hep-ph/9408339 [hep-ph]].

\bibitem{Joyce:1994zn}
M.~Joyce, T.~Prokopec and N.~Turok,
``Nonlocal electroweak baryogenesis. Part 1: Thin wall regime,''
Phys. Rev. D \textbf{53} (1996), 2930-2957
[arXiv:hep-ph/9410281 [hep-ph]];
``Nonlocal electroweak baryogenesis. Part 2: The Classical regime,''
Phys. Rev. D \textbf{53} (1996), 2958-2980
[arXiv:hep-ph/9410282 [hep-ph]].

\bibitem{Kanemura:1993hm}
S.~Kanemura, T.~Kubota and E.~Takasugi,
``Lee-Quigg-Thacker bounds for Higgs boson masses in a two doublet model,''
Phys. Lett. B \textbf{313} (1993), 155-160
[arXiv:hep-ph/9303263 [hep-ph]].

\bibitem{Akeroyd:2000wc}
A.~G.~Akeroyd, A.~Arhrib and E.~M.~Naimi,
``Note on tree level unitarity in the general two Higgs doublet model,''
Phys. Lett. B \textbf{490} (2000), 119-124
[arXiv:hep-ph/0006035 [hep-ph]].

\bibitem{Ginzburg:2005dt}
I.~F.~Ginzburg and I.~P.~Ivanov,
``Tree-level unitarity constraints in the most general 2HDM,''
Phys. Rev. D \textbf{72} (2005), 115010
[arXiv:hep-ph/0508020 [hep-ph]].

\bibitem{Kanemura:2015ska}
S.~Kanemura and K.~Yagyu,
``Unitarity bound in the most general two Higgs doublet model,''
Phys. Lett. B \textbf{751} (2015), 289-296
[arXiv:1509.06060 [hep-ph]].

\bibitem{Nie:1998yn}
S.~Nie and M.~Sher,
``Vacuum stability bounds in the two Higgs doublet model,''
Phys. Lett. B \textbf{449} (1999), 89-92
[arXiv:hep-ph/9811234 [hep-ph]].

\bibitem{Kanemura:1999xf}
S.~Kanemura, T.~Kasai and Y.~Okada,
``Mass bounds of the lightest CP even Higgs boson in the two Higgs doublet model,''
Phys. Lett. B \textbf{471} (1999), 182-190
[arXiv:hep-ph/9903289 [hep-ph]].

\bibitem{Ferreira:2004yd}
P.~M.~Ferreira, R.~Santos and A.~Barroso,
``Stability of the tree-level vacuum in two Higgs doublet models against charge or CP spontaneous violation,''
Phys. Lett. B \textbf{603} (2004), 219-229
[erratum: Phys. Lett. B \textbf{629} (2005), 114-114]
[arXiv:hep-ph/0406231 [hep-ph]].

\bibitem{nEDM:2020crw}
C.~Abel \textit{et al.} [nEDM],
``Measurement of the permanent electric dipole moment of the neutron,''
Phys. Rev. Lett. \textbf{124} (2020) no.8, 081803
[arXiv:2001.11966 [hep-ex]].

\bibitem{ALEPH:2013htx}
G.~Abbiendi \textit{et al.} [ALEPH, DELPHI, L3, OPAL and LEP],
``Search for Charged Higgs bosons: Combined Results Using LEP Data,''
Eur. Phys. J. C \textbf{73} (2013), 2463
[arXiv:1301.6065 [hep-ex]].

\bibitem{ATLAS:2018rvc}
M.~Aaboud \textit{et al.} [ATLAS],
``Search for heavy particles decaying into top-quark pairs using lepton-plus-jets events in proton\textendash{}proton collisions at $\sqrt{s} = 13$   $\text {TeV}$ with the ATLAS detector,''
Eur. Phys. J. C \textbf{78} (2018) no.7, 565
[arXiv:1804.10823 [hep-ex]].

\bibitem{CMS:2019bfg}
A.~M.~Sirunyan \textit{et al.} [CMS],
``Search for charged Higgs bosons in the $H^{\pm}$ $\to$ $\tau^{\pm}\nu_\tau$ decay channel in proton-proton collisions at $\sqrt{s} =$ 13 TeV,''
JHEP \textbf{07} (2019), 142
[arXiv:1903.04560 [hep-ex]].

\bibitem{ATLAS:2020zms}
G.~Aad \textit{et al.} [ATLAS],
``Search for heavy Higgs bosons decaying into two tau leptons with the ATLAS detector using $pp$ collisions at $\sqrt{s}=13$ TeV,''
Phys. Rev. Lett. \textbf{125} (2020) no.5, 051801
[arXiv:2002.12223 [hep-ex]].

\bibitem{ATLAS:2021upq}
G.~Aad \textit{et al.} [ATLAS],
``Search for charged Higgs bosons decaying into a top quark and a bottom quark at $ \sqrt{\mathrm{s}} $ = 13 TeV with the ATLAS detector,''
JHEP \textbf{06} (2021), 145
[arXiv:2102.10076 [hep-ex]].

\bibitem{CMS:2014xfa}
V.~Khachatryan \textit{et al.} [CMS and LHCb],
``Observation of the rare $B^0_s\to\mu^+\mu^-$ decay from the combined analysis of CMS and LHCb data,''
Nature \textbf{522} (2015), 68-72
[arXiv:1411.4413 [hep-ex]].

\bibitem{LHCb:2017rmj}
R.~Aaij \textit{et al.} [LHCb],
``Measurement of the $B^0_s\to\mu^+\mu^-$ branching fraction and effective lifetime and search for $B^0\to\mu^+\mu^-$ decays,''
Phys. Rev. Lett. \textbf{118} (2017) no.19, 191801
[arXiv:1703.05747 [hep-ex]].



\bibitem{Davidson:2005cw}
S.~Davidson and H.~E.~Haber,
``Basis-independent methods for the two-Higgs-doublet model,''
Phys. Rev. D \textbf{72} (2005), 035004
[erratum: Phys. Rev. D \textbf{72} (2005), 099902]
[arXiv:hep-ph/0504050 [hep-ph]].

\bibitem{Glashow:1976nt}
S.~L.~Glashow and S.~Weinberg,
``Natural Conservation Laws for Neutral Currents,''
Phys. Rev. D \textbf{15} (1977), 1958. 

\bibitem{Cabibbo:1963yz}
N.~Cabibbo,
``Unitary Symmetry and Leptonic Decays,''
Phys. Rev. Lett. \textbf{10} (1963), 531-533. 

\bibitem{Kobayashi:1973fv}
M.~Kobayashi and T.~Maskawa,
``CP Violation in the Renormalizable Theory of Weak Interaction,''
Prog. Theor. Phys. \textbf{49} (1973), 652-657. 

\bibitem{Barger:1989fj}
V.~D.~Barger, J.~L.~Hewett and R.~J.~N.~Phillips,
``New Constraints on the Charged Higgs Sector in Two Higgs Doublet Models,''
Phys. Rev. D \textbf{41} (1990), 3421-3441. 

\bibitem{Grossman:1994jb}
Y.~Grossman,
``Phenomenology of models with more than two Higgs doublets,''
Nucl. Phys. B \textbf{426} (1994), 355-384
[arXiv:hep-ph/9401311 [hep-ph]].

\bibitem{Aoki:2009ha}
M.~Aoki, S.~Kanemura, K.~Tsumura and K.~Yagyu,
``Models of Yukawa interaction in the two Higgs doublet model, and their collider phenomenology,''
Phys. Rev. D \textbf{80} (2009), 015017
[arXiv:0902.4665 [hep-ph]].



\bibitem{Akeroyd:2016ymd}
A.~G.~Akeroyd, M.~Aoki, A.~Arhrib, L.~Basso, I.~F.~Ginzburg, R.~Guedes, J.~Hernandez-Sanchez, K.~Huitu, T.~Hurth and M.~Kadastik, \textit{et al.}
``Prospects for charged Higgs searches at the LHC,''
Eur. Phys. J. C \textbf{77} (2017) no.5, 276
[arXiv:1607.01320 [hep-ph]].

\bibitem{Arbey:2017gmh}
A.~Arbey, F.~Mahmoudi, O.~Stal and T.~Stefaniak,
``Status of the Charged Higgs Boson in Two Higgs Doublet Models,''
Eur. Phys. J. C \textbf{78} (2018) no.3, 182
[arXiv:1706.07414 [hep-ph]].

\bibitem{Enomoto:2015wbn}
T.~Enomoto and R.~Watanabe,
``Flavor constraints on the Two Higgs Doublet Models of Z$_{2}$ symmetric and aligned types,''
JHEP \textbf{05} (2016), 002
[arXiv:1511.05066 [hep-ph]].

\bibitem{Haller:2018nnx}
J.~Haller, A.~Hoecker, R.~Kogler, K.~M\"onig, T.~Peiffer and J.~Stelzer,
``Update of the global electroweak fit and constraints on two-Higgs-doublet models,''
Eur. Phys. J. C \textbf{78} (2018) no.8, 675
[arXiv:1803.01853 [hep-ph]].

\bibitem{Jung:2013hka}
M.~Jung and A.~Pich,
``Electric Dipole Moments in Two-Higgs-Doublet Models,''
JHEP \textbf{04} (2014), 076
[arXiv:1308.6283 [hep-ph]].

\bibitem{Cheung:2014oaa}
K.~Cheung, J.~S.~Lee, E.~Senaha and P.~Y.~Tseng,
``Confronting Higgcision with Electric Dipole Moments,''
JHEP \textbf{06} (2014), 149
[arXiv:1403.4775 [hep-ph]].

\bibitem{Barr:1990vd}
S.~M.~Barr and A.~Zee,
``Electric Dipole Moment of the Electron and of the Neutron,''
Phys. Rev. Lett. \textbf{65} (1990), 21-24
[erratum: Phys. Rev. Lett. \textbf{65} (1990), 2920]

\bibitem{Pospelov:2000bw}
M.~Pospelov and A.~Ritz,
``Neutron EDM from electric and chromoelectric dipole moments of quarks,''
Phys. Rev. D \textbf{63} (2001), 073015
[arXiv:hep-ph/0010037 [hep-ph]].

\bibitem{Hisano:2012sc}
J.~Hisano, J.~Y.~Lee, N.~Nagata and Y.~Shimizu,
``Reevaluation of Neutron Electric Dipole Moment with QCD Sum Rules,''
Phys. Rev. D \textbf{85} (2012), 114044
[arXiv:1204.2653 [hep-ph]].

\bibitem{Fuyuto:2013gla}
K.~Fuyuto, J.~Hisano, N.~Nagata and K.~Tsumura,
``QCD Corrections to Quark (Chromo)Electric Dipole Moments in High-scale Supersymmetry,''
JHEP \textbf{12} (2013), 010
[arXiv:1308.6493 [hep-ph]].

\bibitem{Abe:2013qla}
T.~Abe, J.~Hisano, T.~Kitahara and K.~Tobioka,
``Gauge invariant Barr-Zee type contributions to fermionic EDMs in the two-Higgs doublet models,''
JHEP \textbf{01} (2014), 106
[erratum: JHEP \textbf{04} (2016), 161]
[arXiv:1311.4704 [hep-ph]].

\bibitem{Weinberg:1989dx}
S.~Weinberg,
``Larger Higgs Exchange Terms in the Neutron Electric Dipole Moment,''
Phys. Rev. Lett. \textbf{63} (1989), 2333.

\bibitem{Dicus:1989va}
D.~A.~Dicus,
``Neutron Electric Dipole Moment From Charged Higgs Exchange,''
Phys. Rev. D \textbf{41} (1990), 999.

\bibitem{Khatsimovsky:1987fr}
V.~M.~Khatsimovsky, I.~B.~Khriplovich and A.~S.~Yelkhovsky,
``Neutron Electric Dipole Moment, $T$ Odd Nuclear Forces and Nature of {CP} Violation,''
Annals Phys. \textbf{186} (1988), 1-14. 

\bibitem{oblique_parameter}
M.~E.~Peskin and T.~Takeuchi,
``A New constraint on a strongly interacting Higgs sector,''
Phys. Rev. Lett. \textbf{65} (1990), 964-967; 
``Estimation of oblique electroweak corrections,''
Phys. Rev. D \textbf{46} (1992), 381-409. 

\bibitem{Sikivie:1980hm}
P.~Sikivie, L.~Susskind, M.~B.~Voloshin and V.~I.~Zakharov,
``Isospin Breaking in Technicolor Models,''
Nucl. Phys. B \textbf{173} (1980), 189-207.

\bibitem{Haber:1992py}
H.~E.~Haber and A.~Pomarol,
``Constraints from global symmetries on radiative corrections to the Higgs sector,''
Phys. Lett. B \textbf{302} (1993), 435-441
[arXiv:hep-ph/9207267 [hep-ph]].

\bibitem{Pomarol:1993mu}
A.~Pomarol and R.~Vega,
``Constraints on CP violation in the Higgs sector from the rho parameter,''
Nucl. Phys. B \textbf{413} (1994), 3-15
[arXiv:hep-ph/9305272 [hep-ph]].

\bibitem{Gerard:2007kn}
J.~M.~Gerard and M.~Herquet,
``A Twisted custodial symmetry in the two-Higgs-doublet model,''
Phys. Rev. Lett. \textbf{98} (2007), 251802
[arXiv:hep-ph/0703051 [hep-ph]].

\bibitem{Haber:2010bw}
H.~E.~Haber and D.~O'Neil,
``Basis-independent methods for the two-Higgs-doublet model III: The CP-conserving limit, custodial symmetry, and the oblique parameters S, T, U,''
Phys. Rev. D \textbf{83} (2011), 055017
[arXiv:1011.6188 [hep-ph]].

\bibitem{Grzadkowski:2010dj}
B.~Grzadkowski, M.~Maniatis and J.~Wudka,
``The bilinear formalism and the custodial symmetry in the two-Higgs-doublet model,''
JHEP \textbf{11} (2011), 030
[arXiv:1011.5228 [hep-ph]].

\bibitem{Aiko:2020atr}
M.~Aiko and S.~Kanemura,
``New scenario for aligned Higgs couplings originated from the twisted custodial symmetry at high energies,''
JHEP \textbf{02} (2021), 046
[arXiv:2009.04330 [hep-ph]].



\bibitem{Coleman:1973jx}
S.~R.~Coleman and E.~J.~Weinberg,
``Radiative Corrections as the Origin of Spontaneous Symmetry Breaking,''
Phys. Rev. D \textbf{7} (1973), 1888-1910.

\bibitem{Jackiw:1974cv}
R.~Jackiw,
``Functional evaluation of the effective potential,''
Phys. Rev. D \textbf{9} (1974), 1686.

\bibitem{Baum:2020vfl}
S.~Baum, M.~Carena, N.~R.~Shah, C.~E.~M.~Wagner and Y.~Wang,
``Nucleation is More than Critical -- A Case Study of the Electroweak Phase Transition in the NMSSM,''
[arXiv:2009.10743 [hep-ph]].

\bibitem{Dolan:1973qd}
L.~Dolan and R.~Jackiw,
``Symmetry Behavior at Finite Temperature,''
Phys. Rev. D \textbf{9} (1974), 3320-3341.

\bibitem{Parwani:1991gq}
R.~R.~Parwani,
``Resummation in a hot scalar field theory,''
Phys. Rev. D \textbf{45} (1992), 4695
[erratum: Phys. Rev. D \textbf{48} (1993), 5965]
[arXiv:hep-ph/9204216 [hep-ph]].



\bibitem{McLerran:1990de}
L.~D.~McLerran, E.~Mottola and M.~E.~Shaposhnikov,
``Sphalerons and Axion Dynamics in High Temperature {QCD},''
Phys. Rev. D \textbf{43} (1991), 2027-2035.

\bibitem{Giudice:1993bb}
G.~F.~Giudice and M.~E.~Shaposhnikov,
``Strong sphalerons and electroweak baryogenesis,''
Phys. Lett. B \textbf{326} (1994), 118-124
[arXiv:hep-ph/9311367 [hep-ph]].

\bibitem{Moore:2000mx}
G.~D.~Moore,
``Sphaleron rate in the symmetric electroweak phase,''
Phys. Rev. D \textbf{62} (2000), 085011
[arXiv:hep-ph/0001216 [hep-ph]].



\bibitem{Wainwright:2011kj}
C.~L.~Wainwright,
``CosmoTransitions: Computing Cosmological Phase Transition Temperatures and Bubble Profiles with Multiple Fields,''
Comput. Phys. Commun. \textbf{183} (2012), 2006-2013
[arXiv:1109.4189 [hep-ph]].

\bibitem{Cline:2021dkf}
J.~M.~Cline and B.~Laurent,
``Electroweak baryogenesis from light fermion sources: A critical study,''
Phys. Rev. D \textbf{104} (2021) no.8, 083507
[arXiv:2108.04249 [hep-ph]].

\bibitem{Cepeda:2019klc}
M.~Cepeda, S.~Gori, P.~Ilten, M.~Kado, F.~Riva, R.~Abdul Khalek, A.~Aboubrahim, J.~Alimena, S.~Alioli and A.~Alves, \textit{et al.}
``Report from Working Group 2: Higgs Physics at the HL-LHC and HE-LHC,''
CERN Yellow Rep. Monogr. \textbf{7} (2019), 221-584
[arXiv:1902.00134 [hep-ph]].

\bibitem{Bambade:2019fyw}
P.~Bambade, T.~Barklow, T.~Behnke, M.~Berggren, J.~Brau, P.~Burrows, D.~Denisov, A.~Faus-Golfe, B.~Foster and K.~Fujii, \textit{et al.}
``The International Linear Collider: A Global Project,''
[arXiv:1903.01629 [hep-ex]].

\bibitem{CLICdp:2018cto}
P.~N.~Burrows \textit{et al.} [CLICdp and CLIC],
``The Compact Linear Collider (CLIC) - 2018 Summary Report,''
[arXiv:1812.06018 [physics.acc-ph]].



\bibitem{Chung:2009cb}
D.~J.~H.~Chung, B.~Garbrecht, M.~J.~Ramsey-Musolf and S.~Tulin,
``Lepton-mediated electroweak baryogenesis,''
Phys. Rev. D \textbf{81} (2010), 063506
[arXiv:0905.4509 [hep-ph]].

\bibitem{DeVries:2018aul}
J.~De Vries, M.~Postma and J.~van de Vis,
``The role of leptons in electroweak baryogenesis,''
JHEP \textbf{04} (2019), 024
[arXiv:1811.11104 [hep-ph]].

\bibitem{Xie:2020wzn}
K.~P.~Xie,
``Lepton-mediated electroweak baryogenesis, gravitational waves and the $4\tau$ final state at the collider,''
JHEP \textbf{02} (2021), 090
[arXiv:2011.04821 [hep-ph]].

\bibitem{Riotto:1995hh}
A.~Riotto,
``Towards a nonequilibrium quantum field theory approach to electroweak baryogenesis,''
Phys. Rev. D \textbf{53} (1996), 5834-5841
[arXiv:hep-ph/9510271 [hep-ph]].

\bibitem{Riotto:1997vy}
A.~Riotto,
``Supersymmetric electroweak baryogenesis, nonequilibrium field theory and quantum Boltzmann equations,''
Nucl. Phys. B \textbf{518} (1998), 339-360
[arXiv:hep-ph/9712221 [hep-ph]].

\bibitem{Kanemura:2011kx}
S.~Kanemura, K.~Tsumura and H.~Yokoya,
``Multi-tau-lepton signatures at the LHC in the two Higgs doublet model,''
Phys. Rev. D \textbf{85} (2012), 095001
[arXiv:1111.6089 [hep-ph]].

\bibitem{Kanemura:2014dea}
S.~Kanemura, H.~Yokoya and Y.~J.~Zheng,
``Complementarity in direct searches for additional Higgs bosons at the LHC and the International Linear Collider,''
Nucl. Phys. B \textbf{886} (2014), 524-553
[arXiv:1404.5835 [hep-ph]].

\bibitem{Arhrib:2018ewj}
A.~Arhrib, R.~Benbrik, H.~Harouiz, S.~Moretti and A.~Rouchad,
``A Guidebook to Hunting Charged Higgs Bosons at the LHC,''
[arXiv:1810.09106 [hep-ph]].

\bibitem{Belle-II:2018jsg}
E.~Kou \textit{et al.} [Belle-II],
``The Belle II Physics Book,''
PTEP \textbf{2019} (2019) no.12, 123C01
[erratum: PTEP \textbf{2020} (2020) no.2, 029201]
[arXiv:1808.10567 [hep-ex]].

\bibitem{LHCb:2012myk}
R.~Aaij \textit{et al.} [LHCb],
``Implications of LHCb measurements and future prospects,''
Eur. Phys. J. C \textbf{73} (2013) no.4, 2373
[arXiv:1208.3355 [hep-ex]].

\bibitem{Benzke:2010tq}
M.~Benzke, S.~J.~Lee, M.~Neubert and G.~Paz,
``Long-Distance Dominance of the CP Asymmetry in $B\to X_{s,d}+\gamma$ Decays,''
Phys. Rev. Lett. \textbf{106} (2011), 141801
[arXiv:1012.3167 [hep-ph]].

\bibitem{Belle:2018iff}
S.~Watanuki \textit{et al.} [Belle],
``Measurements of isospin asymmetry and difference of direct $CP$ asymmetries in inclusive $B \to X_s \gamma$ decays,''
Phys. Rev. D \textbf{99} (2019) no.3, 032012
[arXiv:1807.04236 [hep-ex]].

\bibitem{Martin:2020lbx}
J.~W.~Martin,
``Current status of neutron electric dipole moment experiments,''
J. Phys. Conf. Ser. \textbf{1643} (2020) no.1, 012002.

\bibitem{Jeans:2018anq}
D.~Jeans and G.~W.~Wilson,
``Measuring the CP state of tau lepton pairs from Higgs decay at the ILC,''
Phys. Rev. D \textbf{98} (2018) no.1, 013007
[arXiv:1804.01241 [hep-ex]].

\bibitem{LISA:2017pwj}
P.~Amaro-Seoane \textit{et al.} [LISA],
``Laser Interferometer Space Antenna,''
[arXiv:1702.00786 [astro-ph.IM]].

\bibitem{Seto:2001qf}
N.~Seto, S.~Kawamura and T.~Nakamura,
``Possibility of direct measurement of the acceleration of the universe using 0.1-Hz band laser interferometer gravitational wave antenna in space,''
Phys. Rev. Lett. \textbf{87} (2001), 221103
[arXiv:astro-ph/0108011 [astro-ph]].

\bibitem{Corbin:2005ny}
V.~Corbin and N.~J.~Cornish,
``Detecting the cosmic gravitational wave background with the big bang observer,''
Class. Quant. Grav. \textbf{23} (2006), 2435-2446
[arXiv:gr-qc/0512039 [gr-qc]].

\bibitem{Enomoto_future}
K.~Enomoto, S.~Kanemura, Y.~Mura, work in progress. 




\end{thebibliography}
\end{document}